\documentclass[%
 reprint,
 amsmath,amssymb,
 aps,
 pre,
]{revtex4-1}

\usepackage{graphicx}
\usepackage{dcolumn}
\usepackage{bm}
\usepackage{amsmath}
\usepackage{subfigure}
\usepackage{epstopdf}
\usepackage{csvsimple,longtable,booktabs}
\usepackage{bm}
\usepackage{array}
\usepackage{tabularx}
\newcolumntype{C}[1]{>{\centering\let\newline\\\arraybackslash\hspace{0pt}}m{#1}}

\usepackage{hyperref}
\usepackage[mathlines]{lineno}

\usepackage{enumerate}

\begin{document}

\preprint{APS/123-QED}

\title{Degree distributions of bipartite networks and their projections}

\author{Demival Vasques Filho}
 \email{d.vasques@auckland.ac.nz}
\author{Dion R.J. O'Neale}%
 \email{d.oneale@auckland.ac.nz}
\affiliation{%
	Department of Physics, University of Auckland\\
	Te P\={u}naha Matatini, University of Auckland\\ 
	Private Bag 92019, Auckland, New Zealand
}%

\date{\today}

\begin{abstract}
Bipartite (two-mode) networks are important in the analysis of social and economic systems as they explicitly show conceptual links between different types of entities. 
However, applications of such networks often work with a projected (one-mode) version of the original bipartite network. The topology of the projected network, and the dynamics that take place on it, are highly dependent on the degree distributions of the two different node types from the original bipartite structure. 
To date, the interaction between the degree distributions of bipartite networks and their one-mode projections is well understood for only a few cases, or for networks that satisfy a restrictive set of assumptions. Here we show a broader analysis in order to fill the gap left by previous studies.  
We use the formalism of generating functions to prove that the degree distributions of both node types in the original bipartite network affect the degree distribution in the projected version. To support our analysis, we simulate several types of synthetic bipartite networks using a configuration model where node degrees are assigned from specific probability distributions, ranging from peaked to heavy tailed distributions.
Our findings show that when projecting a bipartite network onto a particular set of nodes, the degree distribution for the resulting one-mode network follows the distribution of the nodes being projected on to, but only so long as the degree distribution for the opposite set of nodes does not have a heavier tail. 
Furthermore, we show that bipartite degree distributions are not the only feature driving topology formation of projected networks, in contrast to what is commonly described in the literature.

\end{abstract}

\maketitle

\section{\label{sec:introduction}Introduction}
Bipartite structures are of great importance in the analysis of social and economic networks. They can be used to represent conceptual relations --- such as membership, affiliation, collaboration, employment, ownership and others --- between two different types of entities within a system, namely bottom and top nodes, or agents and artifacts \cite{koskinen2012modelling,wasserman1994social,breiger1974duality}. Often, we are particularly interested in one of the types of nodes of a bipartite network (e.g. the agents) and, in order to investigate the relationships among them, we create a new network with only these nodes. This new graph is a projection of the original bipartite network. Connections in this projected network exist only if a pair of nodes share a common neighbor in the original bipartite structure. 

As an example, consider a bipartite network in which executives are connected to companies when they sit on a company's board of directors. From this network, it is possible to construct a projection, which is a network of company directors \cite{ramasco2004Self,newman2001random}, where two agents (i.e. directors) are connected if they sit in the same board. Projections of bipartite networks are frequently used in social and economic systems analysis but, as we will see in Sec. \ref{sec:bipartite} and \ref{sec:arbritary}, there is an inherent loss of information when creating a one-mode network from a bipartite structure \cite{borgatti1997network,zhou2007bipartite}. Moreover, 
the resulting network topology and the dynamics that can take place on a projected network are particularly sensitive to the degree distributions of the underlying bipartite graph \cite{mukherjee2011understanding,nacher2011degree}. Hence, the edges in a projected network are obviously a consequence of the edges between agents and artifacts from the original sets.

In the present paper, we give a more general view of how different degree distributions in bipartite networks affect the distributions of their projections. Studies regarding bipartite structures have shed new light on the topic \cite{newman2001random,guillaume2006bipartite,birmele2009scale,mukherjee2011understanding,nacher2011degree}, however several of these results are applicable for only a few specific cases and with particular assumptions about the degree distributions of the underlying network. 

In \cite{newman2001random} only one projection is built, where the bipartite network has Poisson degree distributions for both node types. Other works \cite{guillaume2006bipartite,birmele2009scale} investigate projections of networks with an exponential degree distribution projected on to nodes with a power law degree degree distribution. In \cite{nacher2011degree} nodes with power law degree distribution are projected onto nodes with power law and exponential degree distributions. Finally, in \cite{mukherjee2011understanding} projections were created using several probability distributions (delta, normal, exponential and power law) projecting onto a $\beta$-distribution \cite{peruani2007emergence,choudhury2010modeling}. Similarly to the latter, we use in this work four probability distributions --- namely delta, Poisson, exponential, and power law distributions --- as the degree distributions of top nodes of the bipartite networks. However, we also use them as degree distributions of bottom nodes. This way it is possible to analyze their combinations and cover a range from very peaked distributions to heavy tailed distributions for both sets of nodes in the network.

The degree distribution of a network is closely linked to its degree-degree correlation, or degree assortativity. Since the degree distribution of a projected network essentially amounts to counting the number of paths of length two for nodes in the bipartite network, it is clear that the projected distribution will be affected by any degree-degree correlation in the bipartite network. While, in some cases, there maybe motivation for considering degree assortative or dissortative bipartite networks (e.g. highly prolific authors may tend to publish papers with large numbers of co-authors), we do not impose any constraints on the degree-degree correlation of the bipartite networks considered here. 

However, we do study the degree assortativity of the one-mode projected networks. It has been observed in a number of cases that a range of different social networks (e.g. \cite{newman2001Structure, grossman1995, watts1998collective, Amaral2000Smallworld, newman2003Function}) but not all (e.g. \cite{newman2002Email}) display degree assortativity, while technological and biological networks (e.g. \cite{watts1998collective, Albert1999www, jeong2001lethality, Martinez1991foodweb}) are typically degree dissortative\cite{newman2002Assortative, newman2003Mixing}. A common feature of those networks displaying degree assortativity is that they are one-mode projections of an underlying bipartite network, while the dissortative technological and biological networks, and the non-assortative social networks, are not. 

It is proposed in \cite{newman2003Social} that the degree assortativity observed in the (projections of bipartite) social networks arises from the community structure of a network, with the agents (bottom nodes) belonging to groups (top nodes). In this case, degree assortativity results in the projected network whenever there is sufficient variation in the degrees of the top nodes --- that is, whenever the top degree distribution is not strongly peaked, relative to the bottom degree distribution --- even without degree-degree correlation in the bipartite network. 

While we are primarily interested in the degree distributions of the projected networks, we take advantage of our configuration model simulations to numerically verify the result from \cite{newman2003Social} for a number of combinations of top and bottom degree distributions. These results, in section \ref{sec:arbritary}, also confirm the observation in \cite{ramasco2004Self} that, for some real-world projected collaboration networks with specific degree distributions, the configuration model is insufficient to reproduce the observed degree assortativity --- specifically in the case where the top degree distribution (e.g. authors per paper) is exponential while the bottom nodes (e.g. papers per author) follow a power law with a high-degree cut-off. 

It is also worth mentioning that, in the previously mention theoretical studies of bipartite networks, the projections are created either as multigraphs, or making the assumption that the underlying networks are sparse enough that there never is more than one common neighbor for any pair of nodes in the bipartite network. The first assumption results in networks that differ from those used in most empirical network analysis, which uses (sometimes weighted) simple graphs (more about projection methods in Sec. \ref{sec:bipartite}). The second assumption does not reflect most real-world networks, in which it is frequent to have pairs of nodes of one type sharing several neighbors of the other type. For instance, it is very likely to find, in a co-authorship network, co-authors of a paper that have many other papers in collaboration. Here, we fill the gap left by previous studies by giving new insights of how the resulting distributions depend on the original networks and on projection methods.

To do so, we use generating functions as a first approach. Many statistical properties, including network observables such as degree distributions, can be derived by using the generating function formalism \cite{newman2001random,wilf2013generatingfunctionology}. A generating function provides a way to represent a sequence as a power series, where the coefficients of the power series are the values of the sequence. An ordinary generating function is given by
\begin{equation}
\label{eq:genfun}	
f(x) = \sum_{n=0}^{\infty}a_{n}x^{n},
\end{equation}
in which the $n$th derivative gives the coefficient $a_{n}$ of the sequence, according to
\begin{equation}
\label{eq:nthderiv}
a_{n} = \dfrac{1}{n!}\dfrac{d^{n}f(x)}{dx^{n}}\Bigr|_{\substack{x=0}}.
\end{equation}

It is important to note that, when using generating functions, we are not concerned with solving the function for $x$. Rather, $x$ is only playing a role of a place holder for the sequence and making it possible to manipulate such sequences as polynomials \cite{pemmaraju2003computational}. 

The outline of the paper is as follows: in Sec. \ref{sec:bipartite} we review basic concepts of bipartite networks. We discuss the role of projections in social networks analysis and the most common methods for creating one-mode projected networks. In Sec. \ref{sec:generating} we revisit generating functions and their applicability in network science. We clarify the methods used to derive expressions for the generating functions for degree distributions of projections. We present new equations for the resulting degree distributions, in order to add new solutions for a mapping of possible outcomes, based on a number of scenarios, for different choices of degree distributions for the bipartite sets. Finally, in Sec. \ref{sec:arbritary}, we create several artificial bipartite networks using the same probability distributions as in the previous section. We assign degrees to the nodes of the bipartite sets by drawing degrees from  probability distributions, using a configuration model. We compare the results from these simulations with those from the generating function approach to explain how the degree distributions of bipartite networks affect the distribution of their projections.

\section{\label{sec:bipartite}Bipartite networks}

 A bipartite network is a graph $B = \{U,V,E\}$, where $U$ and $V$ are disjoint sets of nodes and $E = \{(u,v):u \in U, v \in V\}$ is the set of links connecting nodes. We will refer to the sets $U$ and $V$ as the bottom and top partitions respectively. Nodes in $U$ can only connect to nodes in $V$ and vice-versa. No connections among nodes of the same set are allowed. Each set of nodes can have independent properties, such as the probability distribution of the node degrees, or the number of nodes (system size). Hence, for the sake of notation, we have\\
 $k_{u}$: degree of node $u$ $\in$ $U$;\\
 $d_{v}$: degree of node $v$ $\in$ $V$;\\
 $P_{b}(k)$: degree distribution of bottom nodes in $U$;\\
 $P_{t}(d)$: degree distribution of top nodes in $V$.\\
 \\

 The total number of edges $L$ of the graph is given by
 \begin{equation}
 \label{eq:totalk}	
 |E|=\sum_{u\in U} k_{u} = \sum_{v\in V} d_{v},
 \end{equation}
 
 Many real-world social and economic networks are naturally bipartite structured. 
 Besides the network of company directors given in the introduction, a few other examples are business networks between companies and banks \cite{souma2003complex}, scientific collaborations on publications \cite{newman2001scientific}, football players and clubs they have played for \cite{onody2004complex} and actor-movie networks \cite{watts1998collective}.
 
 All the above examples make use of one of the most interesting properties of a bipartite network, its one-mode projection. 
A projection onto the nodes $U$ (a so-called bottom projection) results in a one-mode network where node $u$ is connected to $u'$, ${u,u'} \in U$, only if there exists a pair of edges $(u,v)$ and $(u',v)$ in $E$ such that $u$ and $u'$ share a common neighbor $v\in V$, in the bipartite graph $B$.  Similarly, in a projection onto $V$ (or top projection) a node $v$ is connected to a node $v'$ in the projection if they share a neighbor $u \in U$.

\begin{figure}[!ht]
	\includegraphics[scale=0.35]{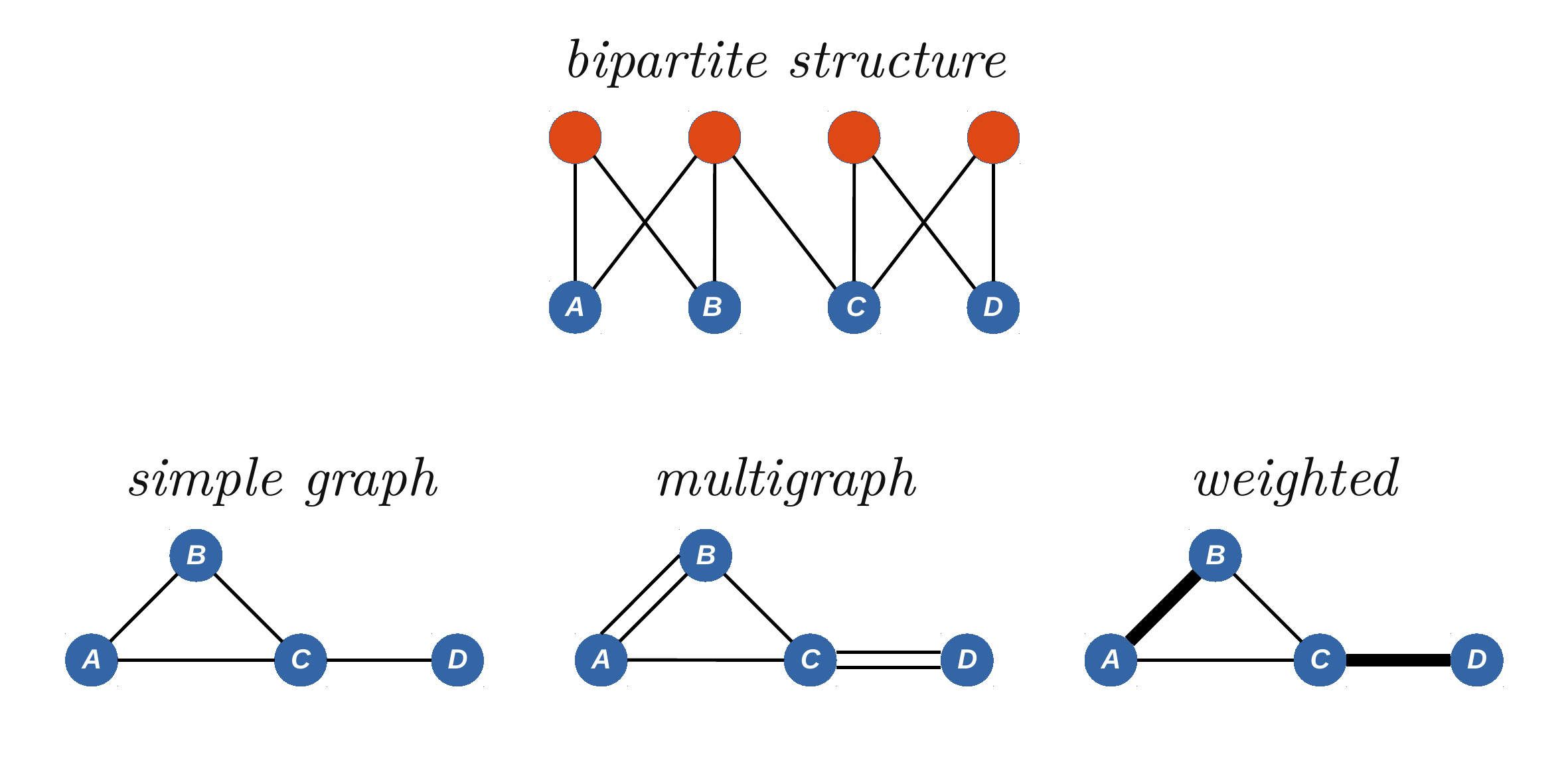}
	\caption{Schematic drawing of three projection methods for bipartite networks: simple graph, multigraph and weighted graph projections}
	\label{fig:projection_methods}
\end{figure}

 Projections can be built using approaches that lead to three different types of graphs: simple graphs, multigraphs and weighted graphs (Fig. \ref{fig:projection_methods}). 
A simple graph is an unweighted graph with no multiple edges between node pairs and no self-loops \cite{gibbons1985algorithmic,west2001introduction}. Hence, in the simple graph projection method, a binary one-mode network will be created. Even though a pair of bottom nodes $u$ and $u'$ can share many common neighbors from the opposite set $V$ in $B$, there will be only a single link connecting such nodes in the projected version. For this case, the number of distinct neighbors of each node in the projection will be equal to the node degree.
 
A multigraph is a graph in which more than one link is allowed between a single pair of nodes \cite{gross2005graph,bollobas2013modern}. 
Then, the multigraph projected network is an unweighted graph in which pairs of nodes can have several links connecting them. In this method, every neighbor $v_{j}$ shared by a pair of nodes $u$ and $u'$ in $B$ results in a link connecting $u$ and $u'$ in the projected graph $G$. Hence, the degree of node $u$ in $G$ can be greater than the number of its distinct neighbors.
 
 Lastly, weighted graphs are networks in which links have values assigned to them \cite{granovetter1973strength,clark1991first,newman2001scientific2}. Such values, that is to say weight, can represent a variety of variables, e.g. duration, intensity, intimacy, exchange of services \cite{granovetter1973strength}, capacity \cite{luczkovich2003defining, pastor2007evolution}, collaboration \cite{newman2001Structure, ramasco2004Self}, among others. In weighted projections, the weight of a edge in the projected network represents the number of common neighbors that $u$ and $u'$ share from the opposite node set $V$ in $B$. The sum of the weights of all links of $u$ is called the node strength of $u$, $s_{u}$ \cite{barrat2004architecture}, given by
\begin{equation}
 s_{u} = \sum_{u'=1}^{|U|} w_{uu'},
 \end{equation}
 where $w_{uu'}$ is the weight of the link connecting nodes $u$ and $u'$ in the weighted projected network. In such a weighted network, the node strength is the same as the node degree in a multigraph projection. 
 
 Weighted projections are commonly used in social network analysis since they allow the ease of working with a simple graph, but still contain information about the number of different connections between a unique node pairs in the projection. The node degree represents the number of neighbors the node has, like in a simple graph projection, while the node strength represents the number of total interactions of each node as the degree in a multigraph projection. For that reason, further in this paper, we will also refer to the node degree of multigraph projections as node strength.
 
 In this work, as we are interested in degree distributions, discussions will focus on simple graphs and multigraphs only. However, it is good to keep in mind that, due to the characteristics discussed above, simple graphs and weighted projections have the same degree distributions when both are built from the same bipartite network. 

 In order to simplify, for the rest of the paper we will refer to projections as bottom projections only, unless otherwise stated. Thus, also for the sake of notation, we have, for the network $G$ formed from the bottom projection of $B$\\
 
 $q_u: \mbox{ degree of node } u\in U, \mbox{ in } G;$\\ 
 $P(q): \mbox{ degree distribution for nodes } u\in U.$\\
 
 A bottom node $u$ $\in$ $U$ now has degree $k_u$ for the bipartite network and $q_u$ for the projected network. At this stage, an important relation can be inferred from such definitions. Considering that node $u$ is connected to $k_{u}$ nodes $v_j\in V$ and each one of these nodes is connected to $d_{v_j}$ nodes $u'\in U$, with $j = {1,2,...,k_{u}}$, then
 
 \begin{equation}
 \label{eq:qi}
 q_{u} \leq \sum_{j=1}^{k_{u}}(d_{v_j} - 1),
 \end{equation}
 where the equality always holds if the projection is a multigraph.
 The degree $q_{u}$ of node $u$ in the bottom projection is the sum of the degrees $d_{v_j}$ of all nodes it is connect to minus its own degree $k_{u}$. When the projection is a (weighted) simple graph, the equality holds only in the case where the bottom nodes never share more than one common neighbor from the top partition. 

Independently of the projection method, there is a inherent loss of information when we create projections of bipartite networks. Starting from an original bipartite structure it is only possible to create a single projected network. However, given a projected network, there is no unique corresponding bipartite structure. Because of the inequality of Eq. \ref{eq:qi} and because of the lack of weight in its links, the simple graph projection is the method with biggest loss of information. We will see more of that in Sec. \ref{sec:arbritary}.

\section{\label{sec:generating}Generating Functions}

``A generating function is a clothesline on which we hang up a sequence of numbers for display''\cite{wilf2013generatingfunctionology}. It is, in fact, just an alternative way of representing any sequence, as a power series notation, whose coefficients are the values of the sequence. 

A simple example is a one-mode network where the degree distribution can be represented using generating functions \cite{newman2001random,li2015generating}. For example, the generating function for the probability of a node in the bottom partition having degree $k$ is given by  
\begin{equation}
\label{eq:probgenfun}	
f_{0}(x) = \sum_{k}P_{b}(k)x^{k}.
\end{equation}
We say that $f_{0}(x)$ is the fundamental generating function for $P_{b}(k)$. Equation (\ref{eq:probgenfun}) is called a probability generating function (PGF); since it describes a probability distribution it must satisfy $f_{0}(1) = 1$, thus
\begin{equation}
f_{0}(1) = \sum_{k}P_{b}(k) = 1.
\end{equation}

An important property of probability generating functions is their moments. If $f_0(x)$ describes the degree distribution of a network, then the first moment of the PGF gives the average degree, $\langle k \rangle$, of the network, according to
\begin{equation}
\label{eq:gfderiv}
f_{0}'(x) = \sum_{k}kP_{b}(k)x^{k-1},
\end{equation}
and
\begin{equation}
\label{eq:avedeg}
f_{0}'(1) = \sum_{k}kP_{b}(k) = \langle k \rangle
\end{equation} 	

Following the approach developed in \cite{newman2001random}, we can use generating functions to find the degree distributions of projected networks. The reasoning is as follows: by selecting a link at random, and following it until reaching a node, we will find that the node has, let's say, degree $k$. The probability $p_{k}$ with which we will reach such a node is proportional to its degree, $p_{k} \propto kP_{b}(k)$. By normalizing, we get
\begin{equation}
p_{k} = \dfrac{kP_{b}(k)}{\sum_{k}kP_{b}(k)} = \dfrac{kP_{b}(k)}{\langle k \rangle}.
\end{equation}
Hence the generating function that corresponds to the probability distribution for a randomly chosen link connecting to a node of degree $k$ is
\begin{equation}
\label{eq:f1x}
f_{1}(x) = {\sum_{k}\dfrac{kP_{b}(k)x^{k-1}}{\langle k \rangle} = \dfrac{f_{0}'(x)}{f_{0}'(1)}}.
\end{equation}

Extending the generating function notation to bipartite networks, we have
\begin{equation}	
f_{0}(x) = \sum_{k}P_{b}(k)x^{k}, \qquad 
g_{0}(x) = \sum_{d}P_{t}(d)x^{d},
\end{equation}	
where, $f_{0}(x)$ and $g_{0}(x)$ are now the generating functions corresponding to $P_{b}(k)$ and $P_{t}(d)$, the degree distributions of the nodes in the partitions $U$ and $V$, respectively. 
Similarly, the average degrees for each partition are given by
$$
f_{0}'(1) = \sum_{k}kP_{b}(k) = \langle{k}\rangle,\quad 
g_{0}'(1) = \sum_{d}dP_{t}(d) = \langle{d}\rangle
$$
and the probability of a randomly chosen link connecting to a node of degree $k$ (respectively, degree $d$) is given by 
$$
f_{1}(x) = \dfrac{f_{0}^{'}(x)}{f_{0}^{'}(1)} = \dfrac{f_{0}^{'}(x)}{\langle k \rangle} \mbox{ and }
g_{1}(x) = \dfrac{g_{0}^{'}(x)}{g_{0}^{'}(1)} = \dfrac{g_{0}^{'}(x)}{\langle d \rangle}.
$$ 

Functions $f_{1}(x)$ and $g_{1}(x)$ are, in fact, representations of the degree distributions of the first neighbors of nodes in $U$ and $V$ respectively. In other words, for instance, $g_{1}(x)$ is the probability distribution of the number of second neighbors of the bottom nodes. As the process of creating a projected network is essentially counting the number of second neighbors of the nodes in the set of interest, the neighborhood of such nodes is a function composition of $f_{0}(x)$ and $g_{1}(x)$. Therefore, the generating function for the degree distribution $P(q)$, of the projection of $B=\{U,V,E\}$ onto $U$, is given by
\begin{equation}
	\label{eq:fog}
	F_{0}(x) = f_{0}(g_{1}(x)) 
	= \sum_{k}P_{b}(k)\left(\dfrac{\sum_{d}dP_{t}(d)x^{d-1}}{\langle{d}\rangle}\right)^{k}.
\end{equation}
This approach assumes that the degree distributions $P_{b}(k)$ and $P_{t}(d)$ in $B$ are independent and the degrees of the bipartite network are uncorrelated. Otherwise, as one can see from Eq. (\ref{eq:fog}), the resulting degree distribution depends on both degree distributions of the bipartite network. This conclusion is also seen in \cite{mukherjee2011understanding}. 

Next, we discuss two different methods for using generating functions to calculate the projected degree distribution of a bipartite network with $P_{b}(k)$ and $P_{t}(d)$ following specific probability distributions.

\subsection{\label{sec:convolution}Convolution Method}

It is possible to derive specific equations for the degree distributions of projected networks as a function of $P_{b}(k)$ and $P_{t}(d)$, the degree distributions of the underlying bipartite networks. The convolution method is one approach to calculating such expressions. 

To represent $P(q)$, we have, from Eq. \ref{eq:fog}
\begin{equation}
\label{eq:Pq}
F_0(x) = \sum_{k}P_{b}(k) \left(\dfrac{\sum_{d}dP_{t}(d)x^{d-1}}{\langle{d}\rangle}\right)^{k}  = \sum_{q}P(q)x^{q}.
\end{equation}

In order to find the coefficients of $x^{q}$ in the right-hand term of the equation, we expand the product:
\begin{multline}
\label{eq:innersum}
\left(\dfrac{\sum_{d}dP_{t}(d)x^{d-1}}{\langle{d}\rangle}\right)^{k} = \\
\left(\sum_{d_{1}} \dfrac{d_{1}P_{t_{d_{1}}}}{\langle{d}\rangle}x^{d_{1}-1}\right)\left(\sum_{d_{2}} \dfrac{d_{2}P_{t_{d_{2}}}}{\langle{d}\rangle}x^{d_{2}-1}\right)...\\
\left(\sum_{d_{k}} \dfrac{d_{k}P_{t_{d_{k}}}}{\langle{d}\rangle}x^{d_{k}-1}\right).
\end{multline}

At this point, one may use the convolution method for the above equation. A convolution of two sequences $\{a_{n}\}_{n=0}^{\infty}$ and $\{b_{m}\}_{m=0}^{\infty}$ is a new sequence $\{c_{k}\}_{k=0}^{\infty}$ \cite{miller2015probability} in which
\begin{equation}
c_{k} = a_{0}b_{k} + a_{1}b_{k-1} + ... + a_{k-1}b_{1} + a_{k}b_{0} = \sum_{l=0}^{k}a_{l}b_{k-l}.
\end{equation}

The states (or place holders) of Eq. (\ref{eq:innersum}) will assume the following form
\begin{equation}
(x^{d_{1}-1})(x^{d_{2}-1})...(x^{d_{k}-1}) = x^{\sum_{j=1}^{k}(d_{j}-1)} = x^{q},
\end{equation}
and we have, therefore
\begin{multline}
\label{eq:pq}
P(q) = \\
\sum_{k}P_{b}(k) \sum_{d_{1} + ... + d_{k} = q + k} \dfrac{d_{1}d_{2}...d_{k}}{\langle d \rangle^{k}}P_{t_{d_{1}}}P_{t_{d_{2}}}...P_{t_{d_{k}}},
\end{multline}
where the relation $d_{1} + d_{2} + ... + d_{k} = q + k$ comes from Eq. (\ref{eq:qi}), assuming that the equality in Eq. (\ref{eq:qi}) holds. This is usually justified in the literature \cite{newman2001random,nacher2011degree} by assuming that the network is sufficiently large that the probability any pair of bottom nodes in the bipartite graph having more than one common neighbor approaches zero. The equality always holds in the case that the projected network is a multigraph.   

$dP_{t}(d)$ is the probability that a node $u \in U$ is connected to a node $v \in V$ with degree $d$. The probability that a node $u \in U$ has degree $k$ in the bipartite and degree $q$ in the projected network can be written as  
\begin{equation}
\label{eq:pkq}
\sum_{d_{1} + d_{2} + ... + d_{k} = q + k} \dfrac{d_{1}d_{2}...d_{k}}{\langle d \rangle^{k}}P_{t_{d_{1}}}P_{t_{d_{2}}}...P_{t_{d_{k}}} = P(k,q).
\end{equation}
An alternate way of representing the degree distribution of bottom projections, using Eqs. (\ref{eq:pq}) and (\ref{eq:pkq}), is
\begin{equation}
\label{eq:pkandpkq}
P(q) = \sum_{k}P_{b}(k)P(k,q).
\end{equation}
This method presents a non-trivial analytical solution. However it is generally difficult to find a closed-form expression for (\ref{eq:pkandpkq}) for specific choices of $P_b(k)$ and $P(k,q)$. Furthermore, the result is limited in its applicability since it requires that the equality in Eq. (\ref{eq:qi}) holds.  

In \cite{mukherjee2011understanding}, the authors use the convolution method assuming that the equality in Eq. (\ref{eq:qi}) holds as they are dealing with multigraph projections. In other words, they are calculating the node strength distribution (multigraph degree distribution) in the resulting projected networks. They solve Eq. (\ref{eq:pkq}) for three specific choices of $P_{t}(d)$ --- delta, normal and exponential distributions --- providing expressions for Eq. (\ref{eq:pkandpkq}), as functions of an arbitrary $P_{b}(k)$ as follows

\begin{equation*}
\ P(q) = 
	\begin{cases}
		P_{b}(k) & \text{if $q=k^{*}(d^{*}-1)$} \\
		0 & \text{otherwise},
	\end{cases} 
\end{equation*}

\begin{equation*}
P(q) = \frac{1}{\sigma \sqrt{2\pi}} \sum_{k} P_{b}(k) k^{-0.5} exp \left(-\frac{(q-k(\langle d \rangle -1))^{2}}{2k\sigma^{2}} \right),
\end{equation*}

\begin{equation*}
P(q) = \frac{1}{a} \sum_{k}P_{b}(k) \frac{exp(-aq)(aq)^{k-1}}{(k-1)!},
\end{equation*}
where $a = 1/(\langle d \rangle - 1)$. The three above expressions for $P(q)$ are for the cases of delta, normal and exponential distributions as $P_{t}(d)$, respectively. Although such expressions provide some insights for the resulting degree distributions $P(q)$, they still are not straightforwardly practical.  

Heavy tail distributions, such as power law distributions, are left aside in \cite{mukherjee2011understanding}. This is because for such distributions the variance and higher order moments can be unbounded, preventing convergence of the convolution sequence.

On the other hand, in \cite{nacher2011degree}, the authors assume a very special case of bipartite networks. In such case, networks are large and sparse such that any pair of nodes $u$ and $u'$ $\in U$ is not allowed to have more than one common neighbor $v \in V$. That is, projections of these networks are simple graphs such that the equality in Eq. (\ref{eq:qi}) always holds. The authors find in \cite{nacher2011degree}, with a slightly different approach, the same Eq. (\ref{eq:pkandpkq}). In contradiction to \cite{mukherjee2011understanding}, they use a power law distribution for $P_{t}(d)$ to perform analysis for the resulting degree distributions for two cases of $P_{b}(k)$ --- exponential and power law distributions. However, to accomplish this, many assumptions and approximations are made, in such a way that, in our understanding, the analytical solutions for both cases become unreliable in practical cases of power law degree distributions. First, such assumptions and approximations are made always considering the presence of nodes with high degree $d$. However, the results presented hold for large absolute values for the exponents of the power law distribution ($\gamma_{t} > 4$), which, in fact, gives rare high degree nodes. Second, first and second moments are not defined in distributions with low exponents -- $\gamma \leq 2$ and $\gamma \leq 3$, respectively. Therefore we don't expect convolutions of sequences such as these to converge.   

\subsection{\label{sec:expansion}Expansion Method}

Another approach to get a closed-form equation, for the degree distribution of the projected network, is by expanding exponential functions. To do so we begin with Eq. (\ref{eq:fog}) (or Eq. (\ref{eq:Pq})) for the degree distribution in the following form
\begin{equation}
F_{0}(x) = \sum_{q}P(q)x^{q}=f_{0}(g_{1}(x))
\end{equation}

From this, we find expressions for $f_{0}(x)$ and $f_{1}(x)$ (or, following our notation, $g_{0}(x)$ and $g_{1}(x)$ in regard to top nodes), for different choices of probability distributions, which can be: delta function, Poisson, exponential, or power law. 

For a delta function distribution, for instance, each node has a degree of exactly $k^*$ and the generating function  has the following form
\begin{equation}
f_{0_{df}}(x)=\sum_{k}P_{b}(k)x^{k}=x^{k^{*}}.
\end{equation}
The generating function for the distribution of remaining links is, from Eq. (\ref{eq:f1x}),
\begin{equation}
\label{eq:gf1df}
f_{1_{df}}(x)=\dfrac{f'_{0_{df}}(x)}{f'_{0_{df}}(1)}=x^{k^{*}-1}.
\end{equation}

Similarly, for the case of a Poisson distribution, in the limit of $N \rightarrow \infty$, we have
\begin{equation}
\label{eq:gfPoisson}
f_{0_{P}}(x)=e^{\langle k \rangle(x-1)}=f_{1_{P}}(x).
\end{equation}

The generating functions for an exponential distribution $f_{0_{exp}}$ and for the remaining links $f_{1_{exp}}$ are given by
\begin{equation}
\label{eq:gfexp}
f_{0_{exp}}(x)=\sum_{k}\lambda e^{-\lambda k} x^{k} = \dfrac{\lambda}{1-xe^{-\lambda}},
\end{equation}
and
\begin{equation}
f_{1_{exp}}(x)=\left(\dfrac{1-e^{-\lambda}}{1-xe^{-\lambda}}\right)^{2},
\end{equation}
where $\lambda=1/\langle k \rangle$.

Finally, for a power law distribution, $P_{b}(k)=\frac{k^{-\gamma}}{\zeta(\gamma)}$, we have

\begin{eqnarray}
f_{0_{pl}}(x)=\dfrac{\mathrm{Li}_{\gamma}(x)}{\zeta(\gamma)}, \\
f_{1_{pl}}=\dfrac{\mathrm{Li}_{\gamma-1}(x)}{x\zeta(\gamma-1)},
\end{eqnarray}
where $\mathrm{Li}_{n}(x)$ is the $n$th polylogarithm of $x$ and $\zeta(s)$ is the Riemann $\zeta$ function.

For a first example, let's say that the degree distribution for the top nodes is a delta function and the degrees of the bottom nodes have a Poisson distribution. In this case $f_{0}(x)$ takes the form of Eq. (\ref{eq:gfPoisson}) and $g_{1}(x)$ takes the form of Eq. (\ref{eq:gf1df}). For the latter, it is worth noticing that now $\langle k \rangle$ has to be replaced by $\langle d \rangle$, as we are dealing with top nodes. Therefore, for this case, the generating function for the bottom projection degree distribution is given by

\begin{equation}
F_{0}(x)=f_{0_{Poisson}}(g_{1_{df}}(x))=e^{\langle k \rangle (x^{d^{*}-1}-1)} .
\end{equation}
By series expansion, one finds that
\begin{equation}
F_{0}(x)=\sum_{q} \frac{\langle k \rangle ^ {\left(\frac{q}{d^{*}-1}\right)} }{e^{\langle k \rangle} \left(\frac{q}{d^{*}-1}\right)!} x^{q},
\end{equation}
and hence, $P(q)$ is given by
\begin{equation}
\ P(q) = 
	\begin{cases}
		\frac{\langle k \rangle ^ {\left(\frac{q}{d^{*}-1}\right)} e^{-\langle k \rangle}}{\left(\frac{q}{d^{*}-1}\right)!} & \text{if $d^{*}-1|q$}\\
		0 & \text{otherwise}.
	\end{cases} 
    \label{eq:deltaPoisson}
\end{equation}

Similarly, for the case where we have delta function distribution for $P_{t}$ and exponential distribution for $P_{b}$, the generating function of $P_{q}$ is
\begin{equation}
F_{0}(x)=f_{0_{exp}}(g_{1_{df}}(x))=\dfrac{\lambda}{1-x^{d^{*}-1}e^{-\lambda}}.
\end{equation}
Again by expanding we have
\begin{equation}
F_{0}(x)=\sum_{q}\lambda (e^{-\lambda})^{q} x^{(d^{*}-1)q},
\end{equation}
and consequently $P(q)$ is given according to
\begin{equation}
\ P(q) = 
	\begin{cases}
		\lambda e^{-\lambda \frac{q}{d^{*}-1}} & \text{if $d^{*}-1|q$}\\
		0 & \text{otherwise}.
	\end{cases} 
    \label{eq:deltaexp}
\end{equation}
Equations \ref{eq:deltaPoisson} and \ref{eq:deltaexp} provide explicit expressions for $P_{b}(k)$ in contrast to the one presented in \cite{mukherjee2011understanding}, using the convolution method, for the cases of a delta function as top degree distributions.

Although delta distributions, either for bottom or top nodes, make it possible to obtain analytic results, and hence gain some insights about the resulting degree distributions, they are unlikely to be found in real-word networks. 

Going further, to a slightly more realistic example, we consider the case where both the bottom and top node degrees are Poisson distributed. In this case, the generating function for the projected degree distribution will assume the following form
\begin{equation}
F_{0}(x)=e^{\langle k \rangle (e^{\langle d \rangle(x-1)}-1)}=e^{-\langle k \rangle} e^{\langle k \rangle e^{-\langle d \rangle} e^{\langle d \rangle x}}.
\end{equation} 
Here, we have the constants $a=\langle k \rangle e^{-\langle d \rangle}$ and $b=\langle d \rangle$. Once more, by expanding the exponential functions, $P(q)$ is semi-explicitly given by
\begin{equation}
P(q)=e^{a - \langle k \rangle}T_{q}(a) \frac{b^{q}}{q!},
\end{equation}
where $T_{q}$ is a Touchard polynomial and $T_{q}(a)$ can be expressed as a polynomial in $a$, using Stirling numbers of the second kind according to
\begin{equation}
T_{q}(a)=\sum^{q}_{n=1}S(q,n)a^{n}.
\end{equation}
This solution was also reasoned in \cite{newman2001random}, presented with a simulation for a network of company directors, briefly discussed earlier in this paper.

A final example, for our matters, is choosing an exponential distribution for bottom nodes and a Poisson distribution for the top set. Then,
\begin{equation}
F_{0}(x)=\dfrac{\lambda}{1-e^{\langle d \rangle (x-1)} e^{-\lambda}} = \dfrac{\lambda}{e^{-\lambda -\langle d \rangle}e^{\langle d \rangle x}}.
\end{equation}
Again, we change our constants making $a=e^{-\lambda -\langle d \rangle}$ and $b=\langle d \rangle$, such that
\begin{equation}
P(q) = \dfrac{ab^{q}}{(1-a)^{(q+1)}q!}A_{q}(a).
\end{equation}
Here, $A_{n}(t)$ denotes the $n$th Eulerian polynomial, which can be defined recursively as $A_{0}(t)=1$ and
\begin{equation}
A_{n}(t)=t(1-t)A'_{n-1}(t)+(1+(n-1)t)A_{n-1}(t).
\end{equation}

At this stage, one may realize that, just like in the convolution method, the expansion method is not suitable for power law distributions. Here, this is mainly due to presence of polylogarithm functions in the solution, and the difficulties of algebraic manipulations of such functions. Once more, we notice that the resulting distributions of projected networks, when the original bipartite sets have degree distributions that follow a power law, have no analytical solutions.

\section{\label{sec:arbritary}Stochastic simulations}
In the previous section we discussed an analytic approach to finding the expected degree distributions of projected networks. 
We saw that, even though generating functions can give us useful insights, few analytical solutions are possible and mostly in restricted scenarios. Significantly, the known analytic solutions do not include the case of bipartite networks with heavy-tailed distributions --- one of the more interesting, and more common, types of distribution found in real-world networks.
For that reason, we make use of a different approach to gain a better understanding of what to expect for degree distributions of projected networks.

In this section, we perform computational simulations by building bipartite networks with specific degree distributions for the bottom and top node sets. These include the cases where the bottom and top sets, $U$ and $V$, have degrees drawn from power law, exponential, Poisson, and delta distributions. 

In order to build such bipartite networks, we use a version of the configuration model \cite{bender1978asymptotic,bollobas1980probabilistic}, in which we create degree sequences for  the bottom and top nodes. We assign to every node $u \in U$, and to every node $v \in V$, a value for its degree, drawn from the chosen distribution for bottom and top node sets. Next, we check that the sum of the node degrees for the top nodes matches that for the bottom nodes; i.e. that Eq. (\ref{eq:totalk}) is respected. In case of $\sum_{i}k_{i}\neq \sum_{j}d_{j}$, we discard a random node from the set with larger degree sum. This process is repeated until Eq. (\ref{eq:totalk}) is satisfied and the total degrees of the top and bottom nodes are equal. Lastly, we add edges to randomly connect the bottom and top nodes, respecting their allocated degrees.

The results presented here are averages over 100 simulated networks for all distribution cases.

When projecting the bipartite networks, we only consider bottom projections. (Results for top projections are identical if the node sets are reversed.) The node sets used in the simulations are of approximately $1000$ nodes. This size is chosen to be commensurate with the size of many real-world bipartite networks in social and economic systems \cite{hidalgo2009building,battiston2004statistical,abbasi2011evolutionary,chakraborty2010weighted}.

The exact size of the node sets $U$ and $V$ depend on the process of discarding nodes to enforce the condition (\ref{eq:totalk}), as detailed above. For example, in the case with power law distributions $P_d \propto d^{-\gamma_t}$ and $P_k \propto k^{-\gamma_b}$ where $\gamma_t$ is larger than $\gamma_b$, the final network can have node sets much smaller than the initial 1000 nodes. In the most extreme cases considered here, the networks with power law exponents of $\gamma_t=5$ and $\gamma_b=2$ have only $\approx 200$ nodes in the bottom distribution since the mean degrees of the top and bottom node sets are $\langle d\rangle\approx 1.2$ and $\langle k \rangle \approx 6.2$. Tables \ref{setsize} and \ref{setsizepl} summarize the average sizes of the node sets and their average degrees over the simulations for different combinations of top and bottom node degree distributions.

\begin{table}[h!]
	\vspace{0.2cm}
	\centering
	\caption[\label{mean}]{Mean of the sizes of sets $U$ and $V$ and their respective mean average degree over 100 runs for each top and bottom distributions. \label{setsize}}
	\vspace{0.2cm}
	\begin{tabular}{C{1.5cm} C{1.5cm} | C{1.2cm} C{1cm} | C{1.2cm} C{1cm}}%
		\toprule
		\bfseries \boldsymbol{$P_{t}(d)$} &  \bfseries \boldsymbol{$P_{b}(k)$}  & \bfseries \boldsymbol{$|V|$} & \bfseries \boldsymbol{$\langle d \rangle$} & \bfseries \boldsymbol{$|U|$} & \bfseries \boldsymbol{$\langle k \rangle$}\\\hline 
		\midrule
		\csvreader[
		before reading={\catcode`\#=12},
		late after line=\\,
		late after last line=\\ \hline,
		after reading={\catcode`\#=6}]%
		{setsizes_avdegree.csv}{1=\one,2=\two,3=\three,4=\four,5=\five,6=\six}{\one & \two &  \three & \four & \five & \six}
	\end{tabular}
\end{table}

\begin{table}[h!]
	\vspace{0.2cm}
	\centering
	\caption[\label{meanpl}]{Mean of the sizes of sets $U$ and $V$ and their respective mean average degree over 100 runs for top and bottom distributions with power law exponents $\gamma_t$ and $\gamma_b$. \label{setsizepl}}
	\vspace{0.2cm}
	\begin{tabular}{C{1.2cm} C{1.2cm} | C{1.2cm} C{1cm} | C{1.2cm} C{1cm}}%
		\toprule
		\bfseries \boldsymbol{$\gamma_{t}$} &  \bfseries \boldsymbol{$\gamma_{b}$}  & \bfseries \boldsymbol{$|V|$} & \bfseries \boldsymbol{$\langle d \rangle$} & \bfseries \boldsymbol{$|U|$} & \bfseries \boldsymbol{$\langle k \rangle$}\\\hline 
		\midrule
		\csvreader[
		before reading={\catcode`\#=12},
		late after line=\\,
		late after last line=\\ \hline,
		after reading={\catcode`\#=6}]%
		{setsizes_avdegree_plonly.csv}{1=\one,2=\two,3=\three,4=\four,5=\five,6=\six}{\one & \two &  \three & \four & \five & \six}
	\end{tabular}
\end{table}

Most analytic results for degree distributions of projected networks assume pairs of nodes from the same node set in the bipartite network do not share any common neighbors. Such an assumption is reasonable for very large, sparse networks. In contrast, the configuration model used for the simulations, allows the bipartite network $B$ to have pairs of nodes in its bottom set $U$ with more than one common neighbor in $V$. The probability of such common neighbors decreases as the sizes of the node sets $U$ and $V$ increase. However, since we deal with finite size networks, shared common neighbors of nodes in $U$ lead to differences between the node degrees of the simple graph compared with the multigraph projection. 

A demonstration of this can be seen in Fig. \ref{fig:fda}, which shows the degree distributions of a simple graph and a multigraph projection, where both projections are created from a bipartite network with bottom and top node sets having a delta-function degree distribution. That is, top nodes and bottom nodes have the same degrees, $d^{*}$ and $k^{*}$, respectively. In this case, the expected degree distribution $P(q)$ is easily calculated using the generating function approach, according to

\begin{equation}
F(x)=x^{k^{*}(d^{*}-1)},
\end{equation}
and therefore, $P(q)$ is given by
\begin{equation*}
\ P(q) = 
	\begin{cases}
		P_{bp_{k^{*}(d^{*}-1)}}=1 & \text{if $q=k^{*}(d^{*}-1)$} \\
		0 & \text{otherwise}.
	\end{cases} 
\end{equation*}

\begin{figure}[!ht]
	\includegraphics[scale=0.50]{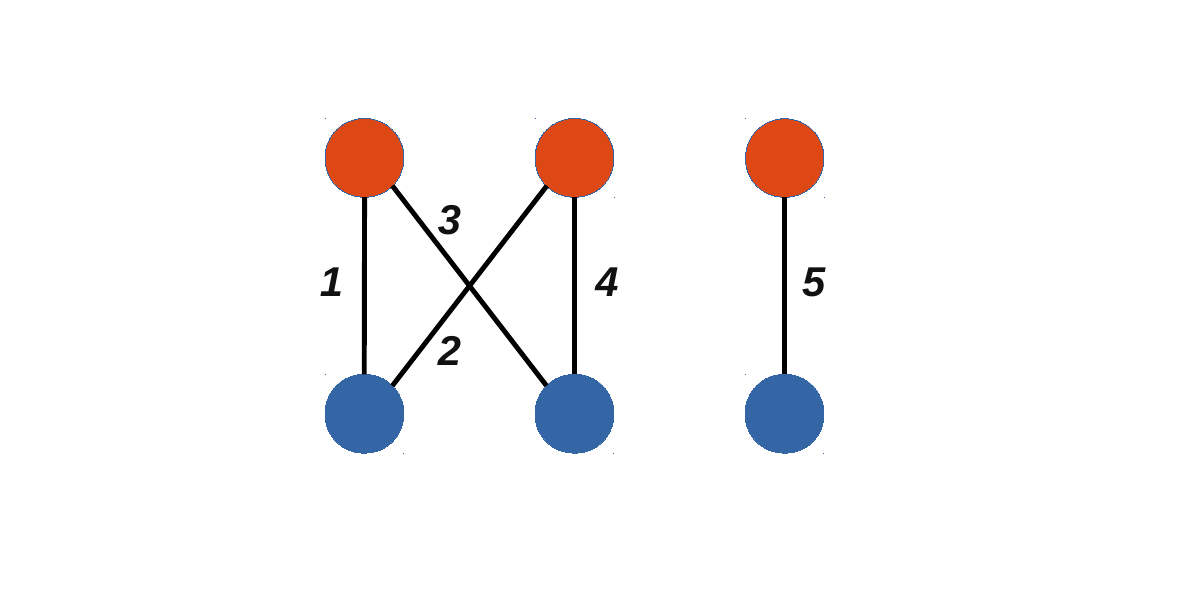}
	\caption{Schematic drawing of a minimal example of the frustration for adding links. Edges have been numbered to indicate the order in which they were added. Although each node has been assigned a target degree of 2, once the first 5 edges have been added, there is no possible way to added the 6th edge without connecting a pair on nodes that are already linked.}
	\label{fig:frustration}
\end{figure}

\begin{figure*}
	\centering
	\subfigure{\label{fig:fda} \includegraphics[scale=0.4]{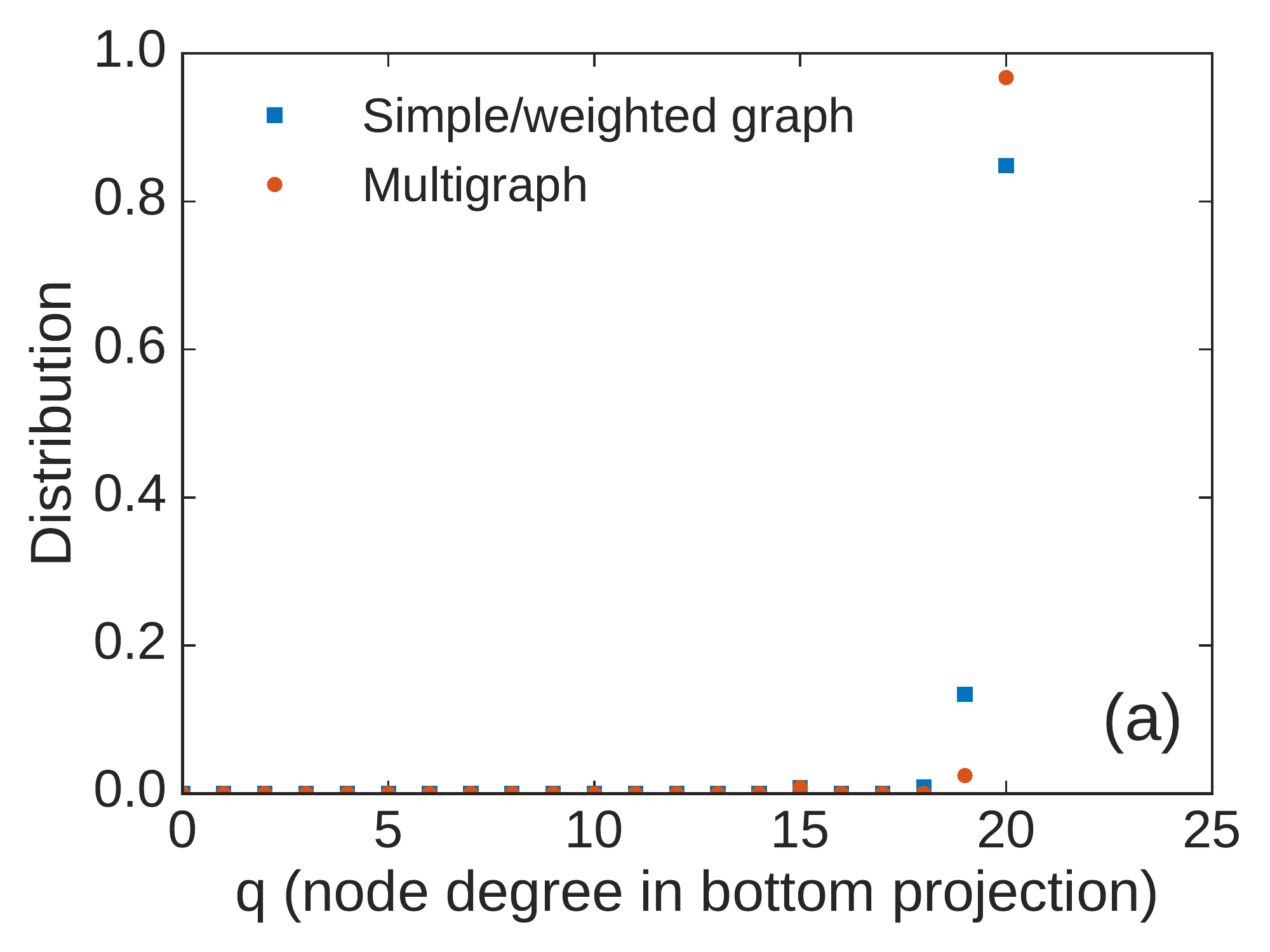}}
	\subfigure{\label{fig:fdb} \includegraphics[scale=0.4]{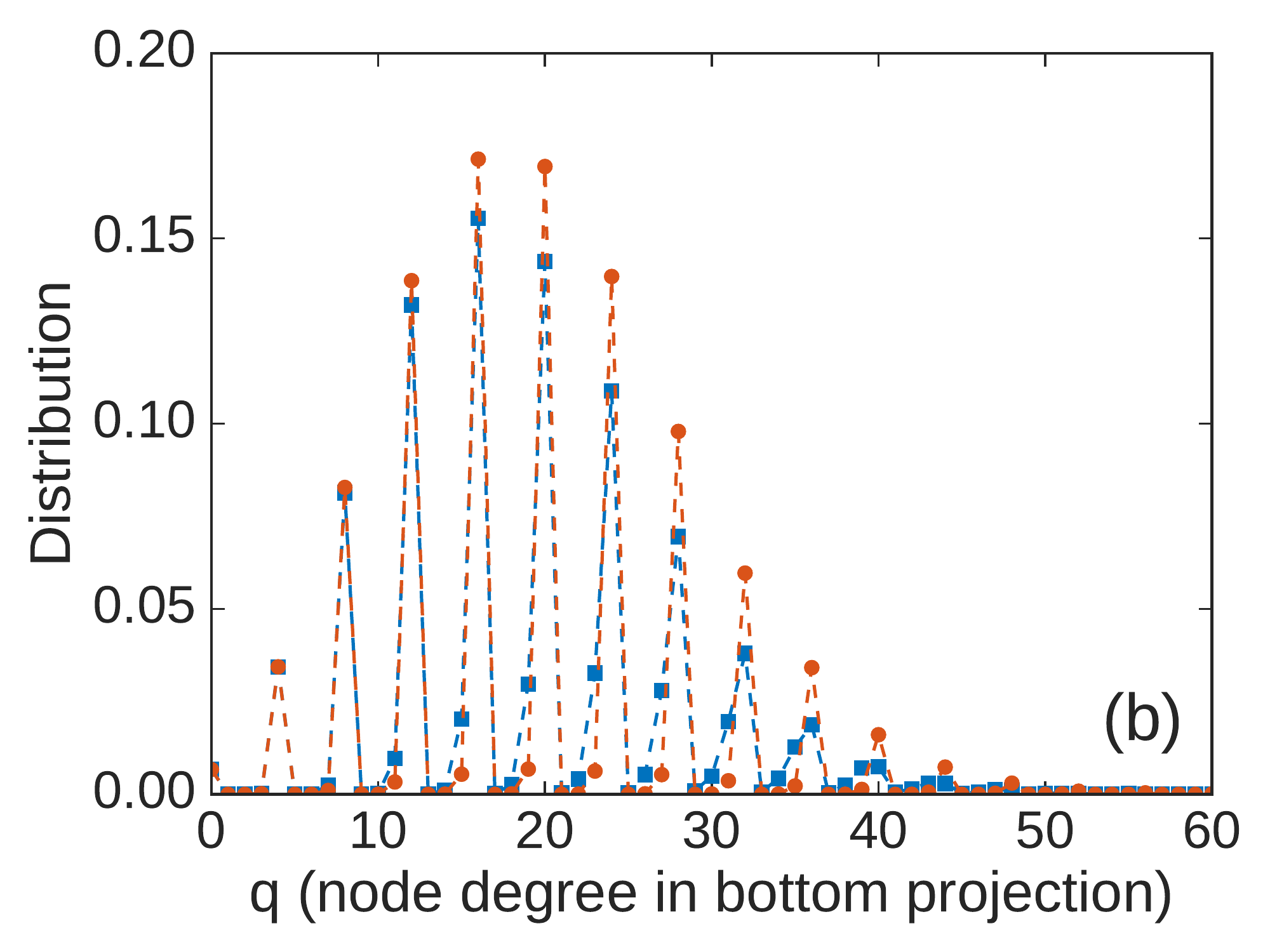}}
	\subfigure{\label{fig:fdc} \includegraphics[scale=0.4]{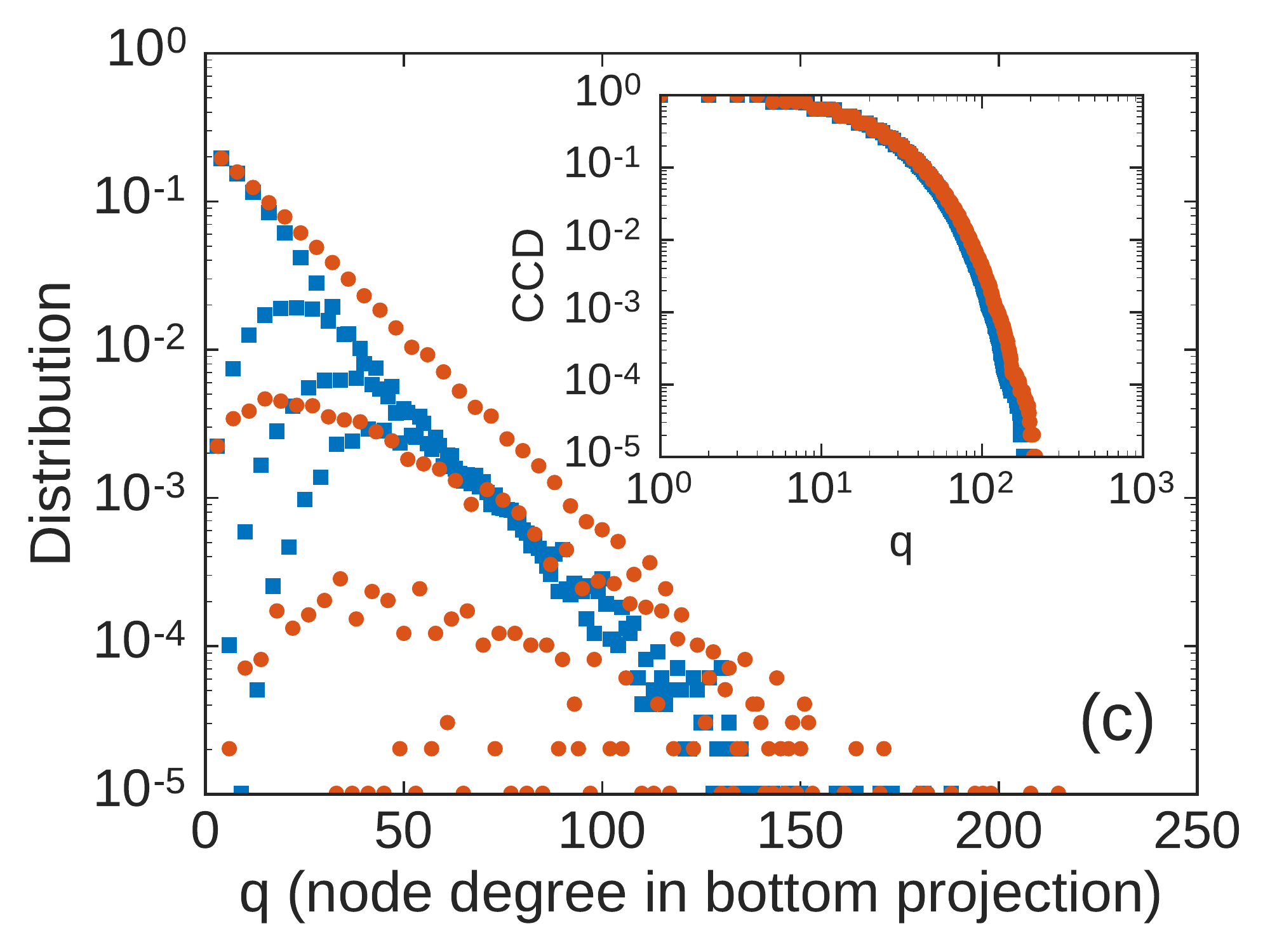}}
	\subfigure{\label{fig:fdd} \includegraphics[scale=0.4]{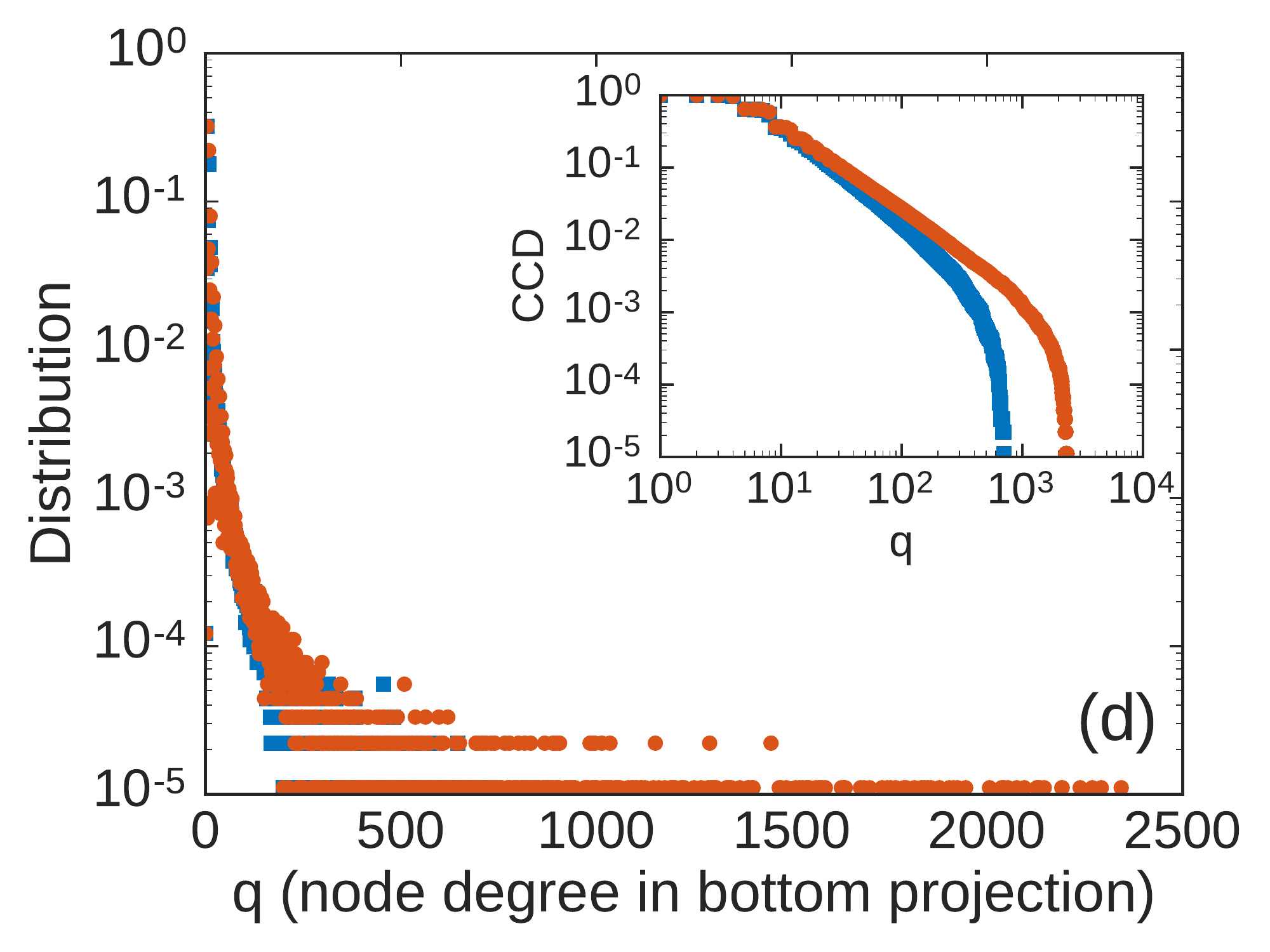}}
	\caption{Degree distributions of projected networks with a delta function as the top distributions. (a) delta function over delta function: top and bottom degrees are  $d^{*}=k^{*}=5$. (b) delta function over Poisson: $d^{*}=\langle k \rangle=5$. (c) delta function over exponential: $d^{*}=\langle k \rangle = 5$. CCD in the inset graph stands for complementary cumulative distribution. (d) delta function over power law: $d^{*}=5$ and $\gamma_{b}=2$ ($\langle k\rangle\simeq5.2$). Similar to the delta over Poisson case, the delta over exponential and delta over power law projected distributions, follow the distribution of their respective bottom degree distributions, modulated by the delta function to give node degrees with multiples of $q=4$, with some broadening of the delta function, due to a small number of nodes with less than their target degree, due to frustration in the edge adding process --- see Fig. \ref{fig:frustration}.}
	\label{fig:fromdelta}
\end{figure*}

Setting $d^{*}=k^{*}=5$, all nodes in the projected network should have $q=20$, following the analytical solution. The multigraph simulation differs from the expected result by only a small amount: $0.02\%$ of nodes in the projection have $q=19$ --- see Fig. \ref{fig:fda}. This is due to the random assignment of links between the two node sets during the construction of the bipartite network. This results in a small fraction of nodes end with $k=4$ (or $d=4$) since, depending on the order in which links are added, it is possible to end up with a bipartite network configuration where it not possible to add an additional edge to a node with degree less than its target degree, without connecting it with a node to which it is already connected. Fig. \ref{fig:frustration} illustrates an example of such a scenario. (We require that the bipartite network be a simple graph since, in general,  real-world examples of bipartite networks have a unique correspondence between pairs of nodes in different sets. For example an author can not appear on the same paper more than once).

In contrast to the multigraph projection, for the simple graph projection, the number of nodes with degree $q=19$ corresponds to about $13\%$ of total nodes. This difference ($13\%$ compared to $0.02\%$) is caused by the presence of common neighbors between pairs of nodes $u$ and $u'$ in $U$, due to the finite size of the network $B$ in our simulations. Multiple links between $u$ and $u'$, in the multigraph projection, are amalgamated into a single link in the simple graph projected network, hence the latter method results in more information loss. 

Similar results are seen for a delta degree distribution of top nodes and Poisson, exponential, or power law degree distributions for bottom nodes (see Figs. \ref{fig:fdb}, \ref{fig:fdc}, and \ref{fig:fdd} respectively); the degree distribution of the projected simple graph is left-shifted, relative to the degree distribution of the multigraph projection, due to shared neighbors of nodes from the bottom node set.

Looking closer at the effects of the different degree distributions on the projected network, we see that the projected degree distributions follow the general form of the degree distributions of the bottom nodes in the bipartite network, but modulated by the delta distribution for the degree of the top nodes. Consequently, the projected degree distribution for a delta  degree distribution over Poisson (Fig. \ref{fig:fdb}), has a projected network with a Poisson degree distribution with spikes at degree $q=d^*-1 =4$, as predicted by Eq. (\ref{eq:deltaPoisson}). The results are similar for the delta degree distribution over exponential (Fig. \ref{fig:fdc}) where the expected degree distribution for the projected network is given by Eq. (\ref{eq:deltaexp}) and for delta distribution over power law --- Fig. \ref{fig:fdd}. In this last case, it is worth noting that power law degree distribution for the projected network does not hold for the tail of the distribution. This is due to finite size effects with the highest degree nodes in the bottom node set being sparsely distributed. Increasing the number of nodes in the network increases the range over which the power law distribution holds for the node degrees of the projected network.

\begin{figure*}
		\centering
		\subfigure{\label{fig:oaa} \includegraphics[scale=0.4]{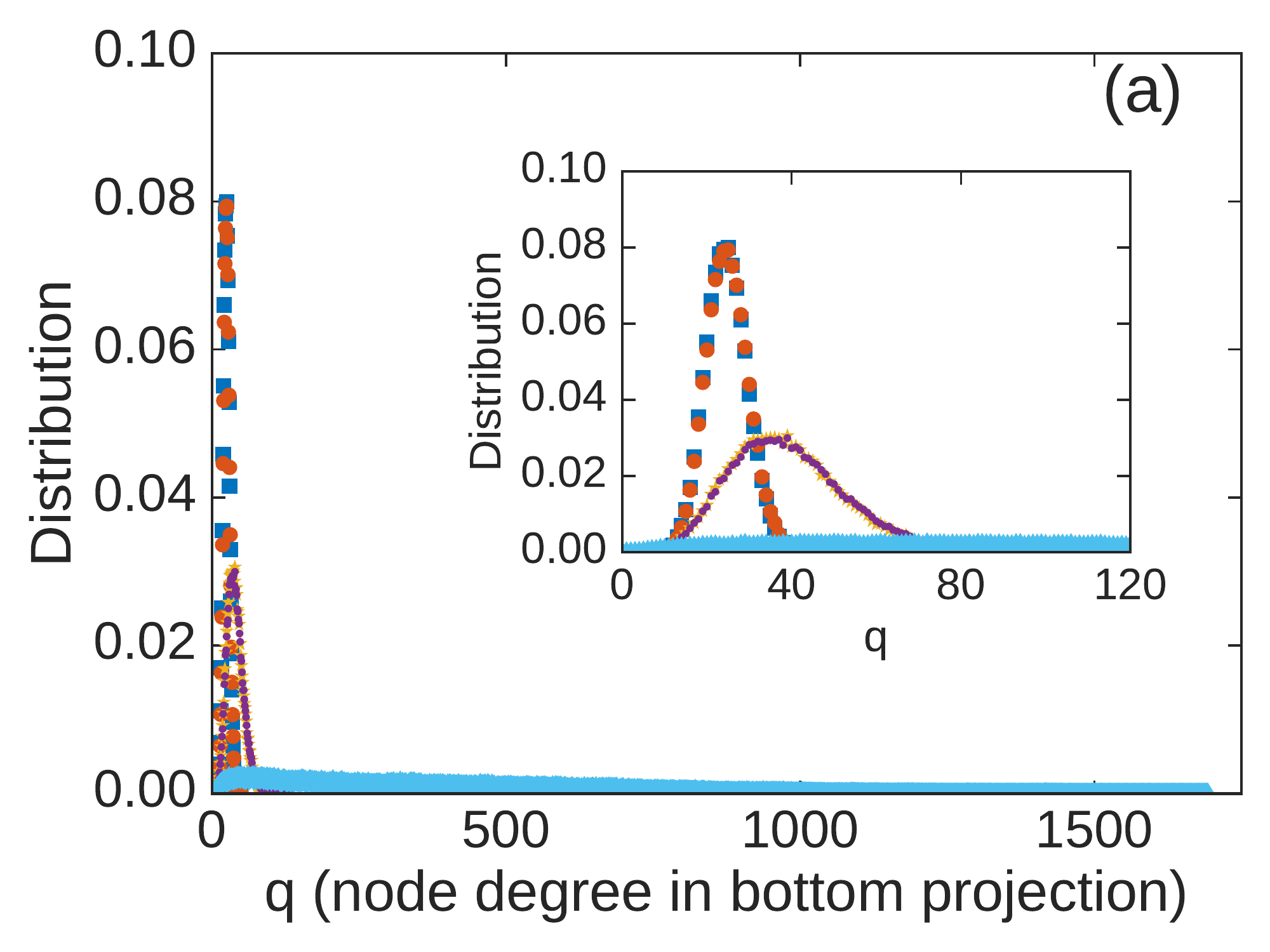}}
		\subfigure{\label{fig:oab} \includegraphics[scale=0.4]{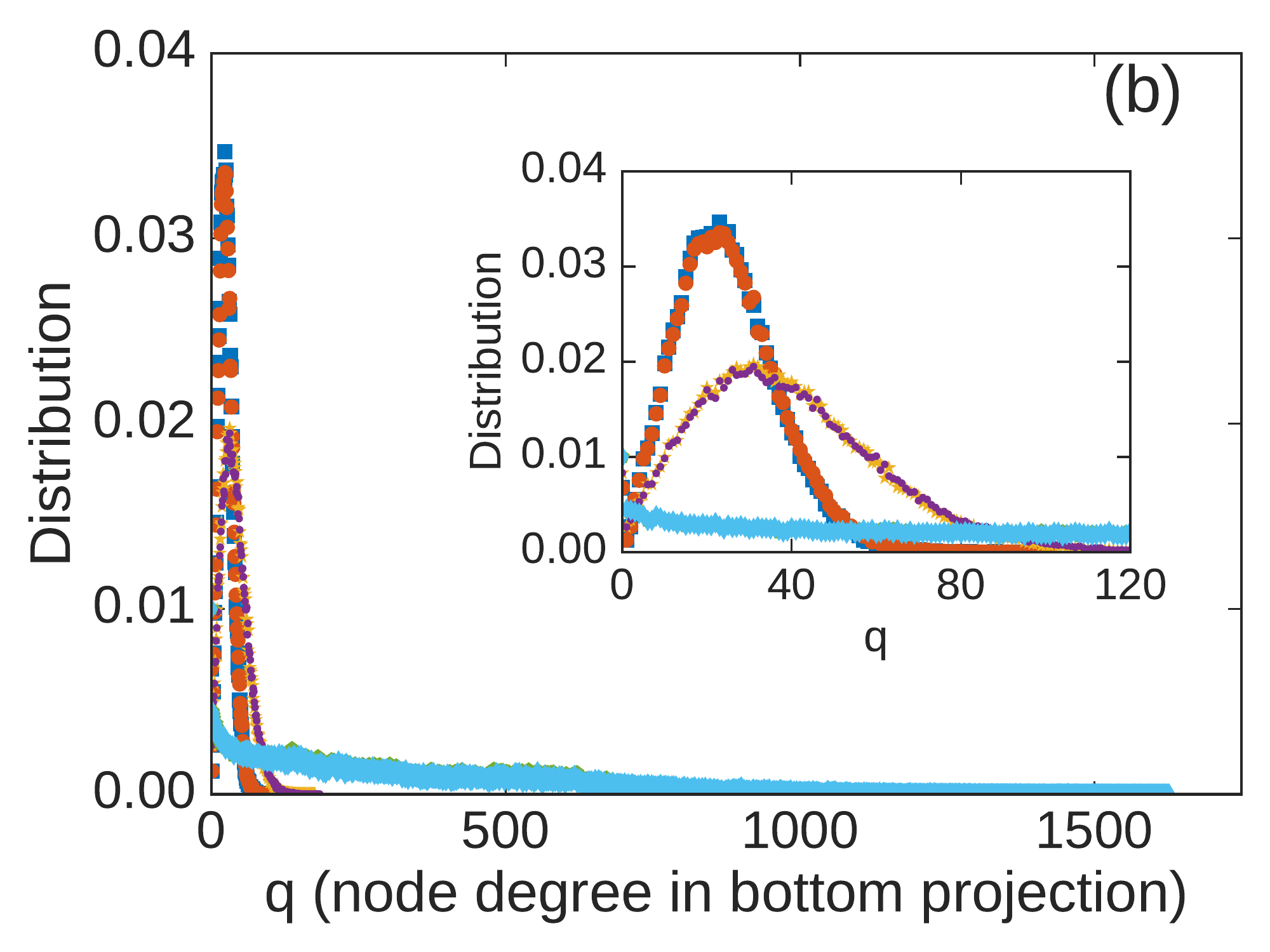}}
		\subfigure{\label{fig:oac} \includegraphics[scale=0.4]{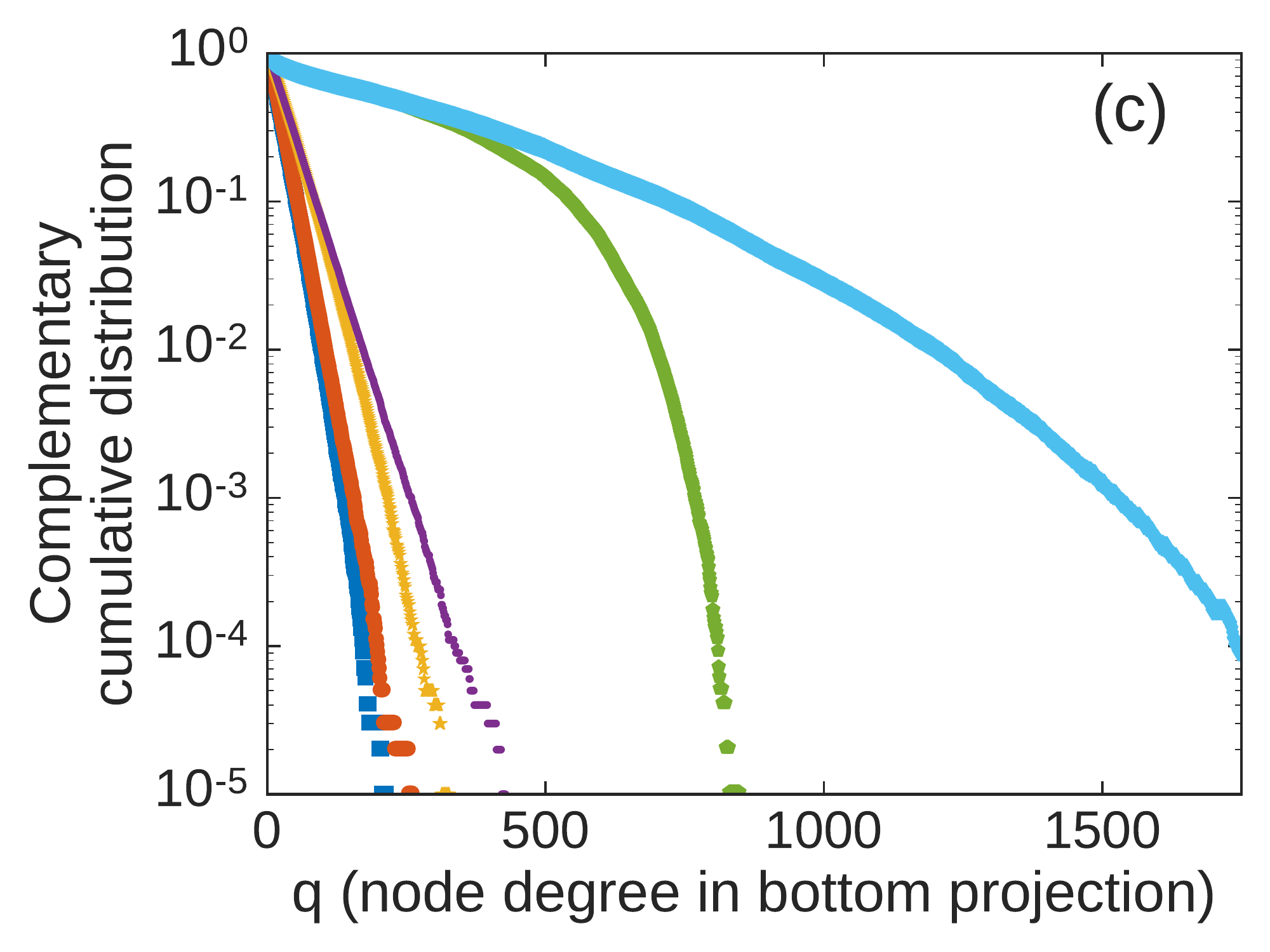}}
		\subfigure{\label{fig:oad} \includegraphics[scale=0.4]{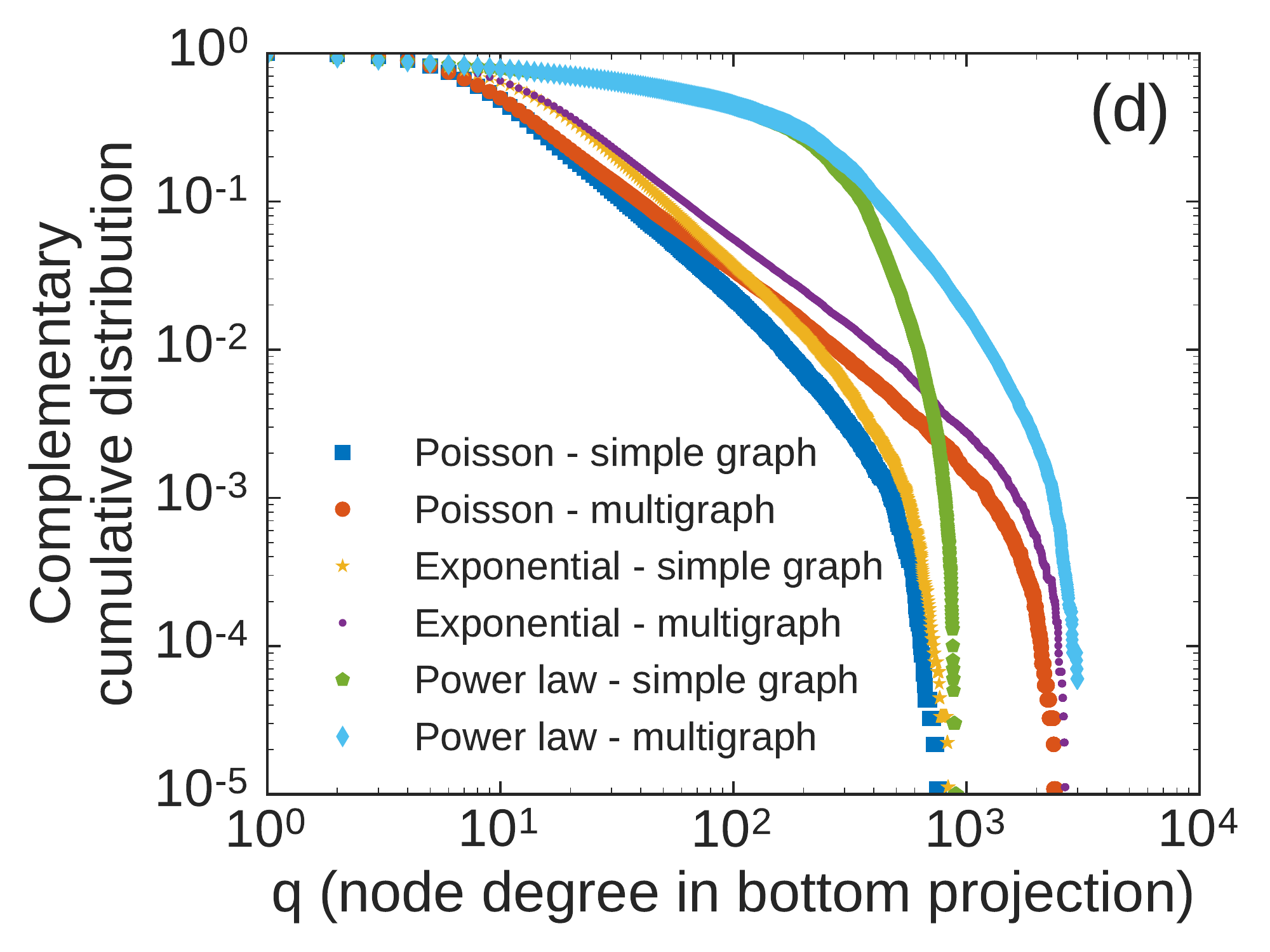}}
		\caption{Degree distributions of projected networks from bipartite networks with a range of top distributions (Poisson: $\langle d\rangle=5$; exponential: $\langle d\rangle=5$; and  power law: $\gamma_t = 2$) and fixed bottom distributions. (a) over delta function: $k^{*}=5$. (b) over Poisson: $\langle k \rangle=5$. (c) over exponential: $\langle k \rangle = 5$. (d) over power law: $\gamma_{b}=2$. The loss of information for the simple graph projections increases, in comparison to the multigraph projections,  as the tail of the bottom distribution becomes heavier. The top distributions also play a role in increasing the loss of information when they are heavy tailed.}
		\label{fig:overall}
	\end{figure*}

More generally, the degree distributions of the projected networks follow the degree distributions of the bottom nodes in the bipartite networks, so long as the degree distributions for the top nodes are more peaked than those of the bottom nodes. 
In cases where the degree distribution of the bottom node set is more peaked than that of the top distribution, the projected network has a degree distribution that is right-shifted and flattened, relative to the top distribution. This can lead to projected networks with extremely heavy-tailed degree distributions, for example the projection of the power law over delta (Fig. \ref{fig:oaa}) network, or power law over Poisson (Fig. \ref{fig:oab}). 

The heuristic that the projected network degree distribution follows that of the bottom nodes in the bipartite network, provided it is more heavy tailed than the top distribution, is well demonstrated by the example of exponential distributed bottom nodes (Fig. \ref{fig:oac}), except for the power law over exponential case. In this latter case the degree distribution for the top nodes has a heavier tail than that of the bottom nodes. As a consequence, the small number of very high degree top nodes are connected to a large fraction of the bottom nodes, leading to high degree cliques in the projected network and a right shifted degree distribution for the projected network. The power law over exponential results in Fig. \ref{fig:oac} also demonstrate again the loss of information for a simple graph projection, relative to the multigraph projection, with the tail of the simple graph projection being left shifted, relative to the multigraph projection. 

The loss of information, due to combining degenerate edges in the simple graph projections, is particularly pronounced for the higher degree nodes in the bottom node set of the bipartite network.
This is because each common neighbor between a pair of bottom nodes in the bipartite network reduces the degree of the same nodes  in the simple graph projection by an amount approximately equal to the original node degrees, relative to a multigraph projection of the same network.
In particular, for heavily right-skewed distributions (e.g. exponential --- Fig. \ref{fig:oac} and power law --- Fig. \ref{fig:oad}) the difference between degree distributions of simple and multigraph projections is also right-skewed, since there is a high probability of shared links between high degree nodes.
This is illustrated, most dramatically, in Fig. \ref{fig:oad} where networks, with power law degree distributions for the bottom nodes, show projections where the degree distributions are influenced by the form of the degree distributions of the top nodes (e.g. Poisson cf. exponential) for the lower degree nodes, while for the higher degree nodes the projected degree distribution is dictated almost entirely by the projection method (simple cf. multigraph projection) and the form of the degree distribution for the top nodes has almost no effect.

We also calculated the degree assortativity $r$ (the degree-degree correlation) of the projected network for each of the combinations of degree distribution types in the bipartite network, according to Eq. (2) of \cite{newman2003Mixing}. Positive values of $r$ indicate networks with degree assortativity (degree-degree correlation), while negative values indicate degree dissortativity (degree-degree anti-correlation). $r=0$ indicates no degree-degree correlation. Our results, presented in table \ref{table:degree_assort_proj} confirm the heuristic presented in \cite{newman2003Social} --- namely that the bottom projection of a bipartite network has degree assortativity when the degree distribution of the top nodes is broader (i.e. more heavy-tailed) than the degree distribution of the bottom nodes. In the inverse case, the projected networks display degree dissortativity. This can be easily understood since top nodes of degree $d$ create cliques, or fully connected sub-graphs of size $d$ in the bottom projection, while bottom nodes of degree $k$ connect together up to $k$ such cliques, in the projection, hence when the top degree distribution is broad, relative to the bottom degree distribution, the bottom projection will be dominated by cliques of nodes with mostly the same degree. In the inverse case, each top node connects a bottom node to a similar number of other bottom nodes, which in turn will have broadly distributed degrees, resulting in degree dissortativity.
 \begin{table}[!h]
 \caption{Mean (standard deviation) values of the degree assortativity of the projected networks with prescribed top (rows) and bottom(columns) degree distributions of the bipartite networks after 100 simulations.}
 \label{table:degree_assort_proj}
 \centering
 \resizebox{\columnwidth}{!}{
 \begin{tabular}{|l|l|l|l|l|l|}
 \hline
             & delta         & Poisson        & exponential    & power law $\gamma=3$    & power law $\gamma=2$    \\
 \hline
 delta       & 0.066 (0.01) & -0.005 (0.009) & -0.019 (0.009) & -0.046 (0.03) & -0.218 (0.04) \\
 Poisson     & 0.193 (0.02) & ~0.026 (0.009)  & -0.018 (0.008) & -0.038 (0.03) & -0.230 (0.03)  \\
 exponential & 0.250 (0.04)  & ~0.090 (0.03)   & -0.014 (0.008) & -0.033 (0.04) & -0.237 (0.03) \\
 power law $\gamma=3$ & 0.470 (0.1)  & ~0.396 (0.1)  & ~0.178 (0.1)  & ~0.302 (0.1)  & -0.219 (0.1) \\
 power law $\gamma=2$ & 0.228 (0.1) & ~0.133 (0.09)  & ~0.009 (0.06)  & ~0.200 (0.2)    & -0.209 (0.06) \\
 \hline
 \end{tabular}
 }
 \end{table}

The case of power law degree distributions for both the bottom and top node sets is worth further investigation, in the context of our claim that the bottom degree distribution dictates the projected degree distribution, as long as it is heavier tailed than the top distribution.

\begin{figure*}
		\centering
		\subfigure{\label{fig:smt2} \includegraphics[scale=0.35]{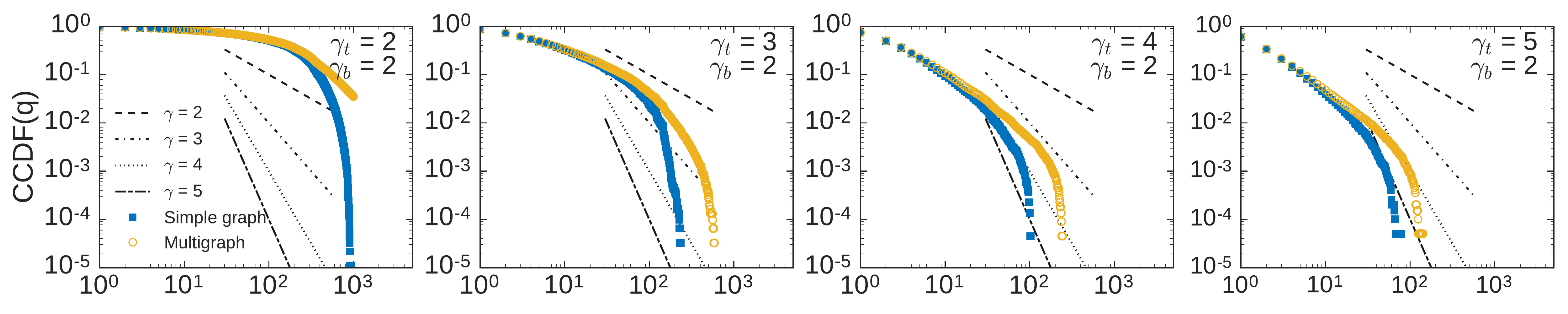}}
		\subfigure{\label{fig:smt3} \includegraphics[scale=0.35]{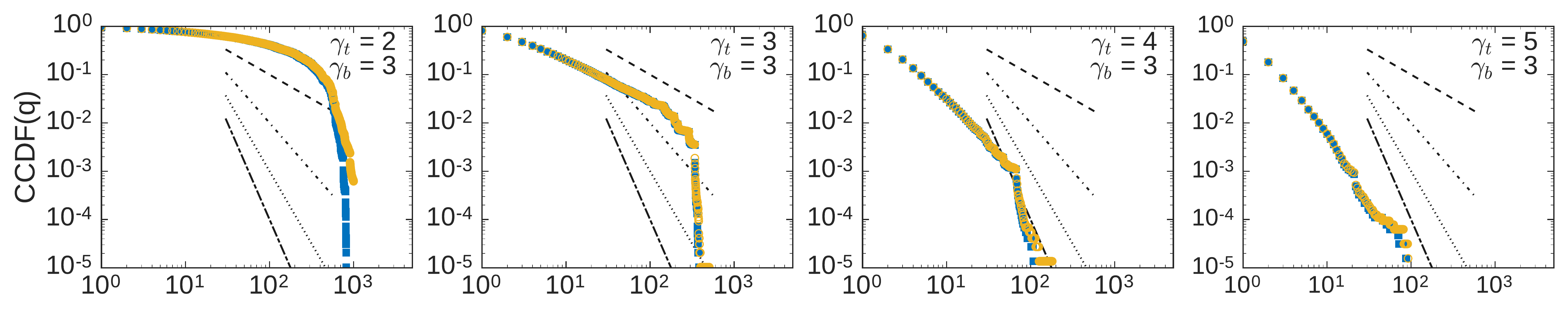}}
		\subfigure{\label{fig:smt4} \includegraphics[scale=0.35]{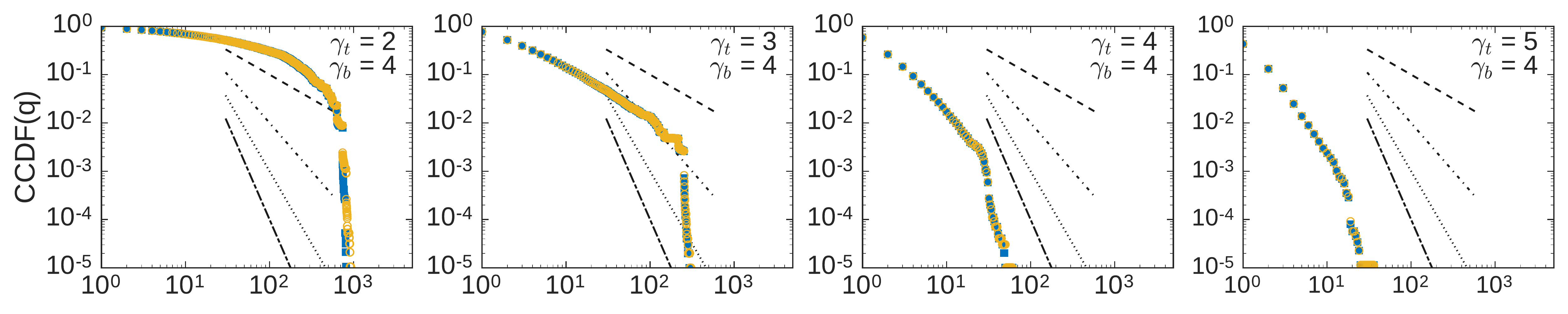}}
		\subfigure{\label{fig:smt5} \includegraphics[scale=0.35]{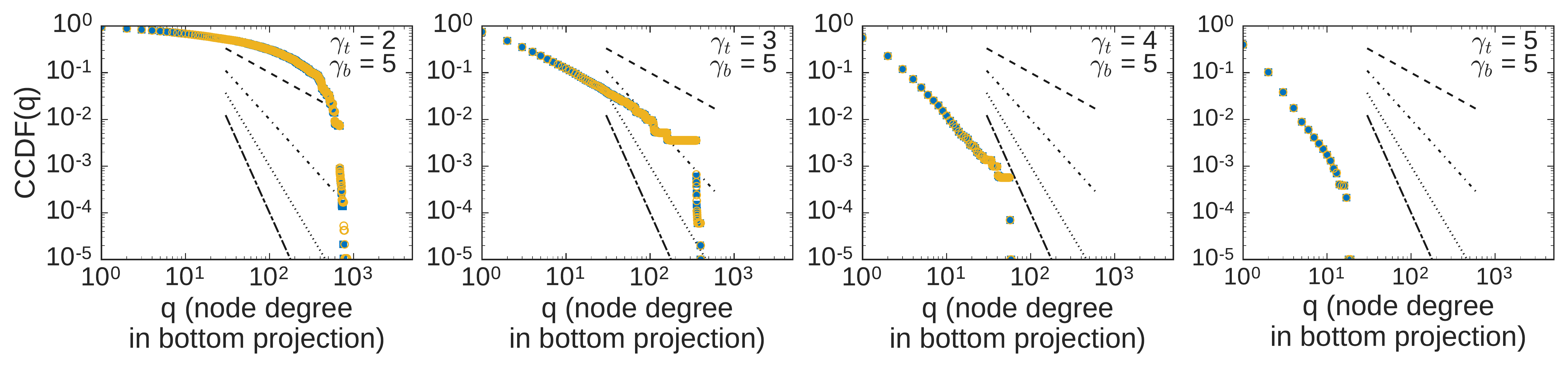}}
		\caption{Degree distributions of projected networks from bipartite networks with power law distribution for bottom and top nodes. Theoretical power law distributions are also shown. Projections were created using simple graphs and multigraphs in all four cases. In the plots above the diagonal ($\gamma_t > \gamma_b$), the projected degree distributions follow the degree distribution of the bottom node set. For the case of $\gamma_t \leq \gamma_b$, where the bottom degree distribution is no longer more heavy tailed than the top, the distributions show the same flattening of the distribution that was observed for the results in Fig. \ref{fig:overall}. This is due to large cliques in the projected network. The turning point in the distributions is due to the finite size of the node sets. Increasing the size of the network causes the turning point to be shifted to the right.}
		\label{fig:simplemultitheoretical}
	\end{figure*}

We therefore take a closer look at bipartite networks with power law distributions in both sets $U$ and $V$. We assign $\gamma = 2, 3, 4$ and $5$, as exponents for the distributions, i.e. $P_{k}\propto k^{-\gamma_{b}}$ and $P_{d}\propto d^{-\gamma_{t}}$, for bottom and top nodes respectively. We interchange them in sets $U$ and $V$, in order to get all 16 possible combinations. Projected networks created using simple graphs and multigraphs are shown in Fig. \ref{fig:simplemultitheoretical} along with the slopes of theoretical power law distributions --- represented by the dashed and dot-dashed lines.

We observe that for those plots above the diagonal (i.e. $\gamma_t > \gamma_b$) in Fig. \ref{fig:simplemultitheoretical}, the projected degree distributions do indeed follow the degree distribution of the bottom node set, including following $\gamma_b$, the expected slope of the bottom degree distribution. For the case of $\gamma_t \leq \gamma_b$, where the bottom degree distribution is no longer more heavy tailed than the top, the distributions for the projected networks remain power law like (we simply observe that they have a linear for a significant range when plotted on log-log axes --- we do not attempt to rule out other possible distributions that may fit equally well). 
However, they show the same flattening of the distribution that was observed for the results in Fig. \ref{fig:overall}, due to large cliques in the projected network, formed by the relative abundance of high degree nodes in the top node set. This flattening increases the more heavy tailed the top distribution is, relative to the bottom. As a rule-of-thumb, if $\gamma_t \leq \gamma_b$ then the slope of the distribution for the projected network is roughly that which corresponds to $\gamma = \gamma_t -1$.

It is also worth noting that the cumulative degree distributions in Fig. \ref{fig:simplemultitheoretical} all display a turning point where the slope of the distribution changes for low degree nodes compared with that for high degree nodes. As mentioned previously, this is a finite size effect, due to the stochastic model sampling the sparse tail of the power law distributions. Increasing the size of the node sets in the simulations, moves this turning point to the right and extends the range over which the projected network follows the predicted degree distribution.

\section{\label{sec:conclusions}Conclusions}

In this paper, we have studied how the degree distributions of bipartite networks affect the degree distributions of the projected (one-mode) networks that can be produced from them. 
To do so, we considered bipartite networks with sets of nodes presenting a range probability distributions. We also considered the effects of different projection approaches, namely simple graph versus multigraph projections, and the consequences of dealing with finite size networks where pairs of nodes from one node set are able to share common neighbors from the opposite node set.

Our initial approach was to use generating functions in order to reach any analytic predicts of how node degrees in the projected networks are distributed. Top ($P_{t}(d)$) and bottom ($P_{b}(k)$) degree distributions were chosen arbitrarily, ranging from peaked to heavy tailed distributions (delta-function, Poisson, exponential, and power law probability distributions). We discussed two ways to solve these functions, namely convolution and expansion methods. While both these methods can be useful for predicting the resulting degree distribution, $P(q)$, of a bottom projection, the analytic approach comes with some caveats. First, generating functions are only suitable for multigraph projections, since each edge in the bipartite graph contributes to the node degrees in the projected network. This is not the case for simple graph projections when pairs of nodes from the bottom node set are allowed to have common neighbors in the opposite node set --- a case which is likely to occur for networks of finite size.
Second, not all probability distributions are amenable to deriving an analytical solution for the projected degree distribution. Cases where degree distributions are heavy tailed --- a kind of distribution often studied in real-world networks --- are hard to solve analytically.

Due to these two aforementioned predicaments of generating functions, we also performed computational simulations as our second approach to reach our goals. Then, we could fill the gaps left by the generating functions approach, and also obtain new findings of what to expect of the structure of projected networks, by taking in account the original bipartite structure. 

The simulations show that if the projection of the bipartite network is onto the bottom node set $U$, then the resulting projected degree distribution, $P(q)$, tends to follow the distribution of the bottom node set, $P_{b}(k)$, so long as the degree distribution of the bottom node set is more right-skewed, or heavy tailed, than the top distribution. In the case that this does not hold, $P(q)$ is subject to the tail of the degree distribution, $P_{t}(d)$, of the top node set, which leads to a flattening of the degree distribution for the projection, relative to that for the bottom node set.
The reason for this flattening is the relative abundance of nodes with high degree in the top distribution. Such nodes induce large cliques (complete subgraphs) in the projected network. Therefore, the projection will have many highly connected nodes. This phenomenon is illustrated particularly well by the case where both top and bottom node sets follow a power law degree distribution, but where the exponent for the top distribution is lower (more heavy tailed) than the bottom. The simulations also reveal a finite size effect, whereby the degree distribution of the projected network doesn't follow the predicted degree distribution for the largest node sizes when one of the node sets in the bipartite network is power law distributed. This leads to turning points in the degree distributions of the projected networks when data is sparse, towards the extreme values of heavy tailed distributions. This result has implications for real-world bipartite networks, which can commonly be of sizes comparable to those simulated here. It suggests that one is unlikely to find fully power law degree distributions for projected networks, even if the statistics of one of the node sets in the bipartite network, do follow a power law.

Lastly, there is an inherent loss of information when projections of bipartite networks are created. A consequence of this is that there is not an unique solution for rebuilding a bipartite network from its projection. Such loss of information is accentuated in simple graph projections where multiple links between pairs of nodes are amalgamated into a single link. 
This divergence between the simple and multigraph degree distributions for the projected network becomes more prevalent, the heavier the tail of the original degree distributions.
When performing social network analysis one must be aware of these differences and understand which characteristics of each projection method best contribute to its studies.

Although non-trivial, the degree distribution of a projected network is one of the more superficial network features. Network features such as clustering, communities, component size distributions, motifs, and a variety of centrality measures may influence the behavior of, and on, networks in different applications. While several methods to infer bipartite network structure have been proposed --- such as modularity  \cite{newman2006finding,barber2007modularity}, motif-based clustering and community structure \cite{zhang2008clustering} or bipartite stochastic block models \cite{larremore2014efficiently} ---, further work is necessary to understand how they arise and interact in the context of bipartite networks and their projections.

\begin{acknowledgments}
The authors would like to thank Jurij Vol\u{c}i\u{c} and Mark Wilson for helpful conversations. 
This work was supported by funding from Callaghan Innovation, New Zealand.
\end{acknowledgments}

\bibliography{degreedistributionsNotes} 

\begin{thebibliography}{52}%
\makeatletter
\providecommand \@ifxundefined [1]{%
 \@ifx{#1\undefined}
}%
\providecommand \@ifnum [1]{%
 \ifnum #1\expandafter \@firstoftwo
 \else \expandafter \@secondoftwo
 \fi
}%
\providecommand \@ifx [1]{%
 \ifx #1\expandafter \@firstoftwo
 \else \expandafter \@secondoftwo
 \fi
}%
\providecommand \natexlab [1]{#1}%
\providecommand \enquote  [1]{``#1''}%
\providecommand \bibnamefont  [1]{#1}%
\providecommand \bibfnamefont [1]{#1}%
\providecommand \citenamefont [1]{#1}%
\providecommand \href@noop [0]{\@secondoftwo}%
\providecommand \href [0]{\begingroup \@sanitize@url \@href}%
\providecommand \@href[1]{\@@startlink{#1}\@@href}%
\providecommand \@@href[1]{\endgroup#1\@@endlink}%
\providecommand \@sanitize@url [0]{\catcode `\\12\catcode `\$12\catcode
  `\&12\catcode `\#12\catcode `\^12\catcode `\_12\catcode `\%12\relax}%
\providecommand \@@startlink[1]{}%
\providecommand \@@endlink[0]{}%
\providecommand \url  [0]{\begingroup\@sanitize@url \@url }%
\providecommand \@url [1]{\endgroup\@href {#1}{\urlprefix }}%
\providecommand \urlprefix  [0]{URL }%
\providecommand \Eprint [0]{\href }%
\providecommand \doibase [0]{http://dx.doi.org/}%
\providecommand \selectlanguage [0]{\@gobble}%
\providecommand \bibinfo  [0]{\@secondoftwo}%
\providecommand \bibfield  [0]{\@secondoftwo}%
\providecommand \translation [1]{[#1]}%
\providecommand \BibitemOpen [0]{}%
\providecommand \bibitemStop [0]{}%
\providecommand \bibitemNoStop [0]{.\EOS\space}%
\providecommand \EOS [0]{\spacefactor3000\relax}%
\providecommand \BibitemShut  [1]{\csname bibitem#1\endcsname}%
\let\auto@bib@innerbib\@empty
\bibitem [{\citenamefont {Koskinen}\ and\ \citenamefont
  {Edling}(2012)}]{koskinen2012modelling}%
  \BibitemOpen
  \bibfield  {author} {\bibinfo {author} {\bibfnamefont {J.}~\bibnamefont
  {Koskinen}}\ and\ \bibinfo {author} {\bibfnamefont {C.}~\bibnamefont
  {Edling}},\ }\href@noop {} {\bibfield  {journal} {\bibinfo  {journal} {Soc.
  Networks}\ }\textbf {\bibinfo {volume} {34}},\ \bibinfo {pages} {309}
  (\bibinfo {year} {2012})}\BibitemShut {NoStop}%
\bibitem [{\citenamefont {Wasserman}\ and\ \citenamefont
  {Faust}(1994)}]{wasserman1994social}%
  \BibitemOpen
  \bibfield  {author} {\bibinfo {author} {\bibfnamefont {S.}~\bibnamefont
  {Wasserman}}\ and\ \bibinfo {author} {\bibfnamefont {K.}~\bibnamefont
  {Faust}},\ }\href@noop {} {\emph {\bibinfo {title} {Social Network Analysis:
  Methods and Applications}}},\ Vol.~\bibinfo {volume} {8}\ (\bibinfo
  {publisher} {Cambridge University Press},\ \bibinfo {address} {Cambridge},\
  \bibinfo {year} {1994})\BibitemShut {NoStop}%
\bibitem [{\citenamefont {Breiger}(1974)}]{breiger1974duality}%
  \BibitemOpen
  \bibfield  {author} {\bibinfo {author} {\bibfnamefont {R.}~\bibnamefont
  {Breiger}},\ }\href@noop {} {\bibfield  {journal} {\bibinfo  {journal} {Soc.
  Forces}\ }\textbf {\bibinfo {volume} {53}},\ \bibinfo {pages} {181} (\bibinfo
  {year} {1974})}\BibitemShut {NoStop}%
\bibitem [{\citenamefont {Ramasco}\ \emph {et~al.}(2004)\citenamefont
  {Ramasco}, \citenamefont {Dorogovtsev},\ and\ \citenamefont
  {Pastor-Satorras}}]{ramasco2004Self}%
  \BibitemOpen
  \bibfield  {author} {\bibinfo {author} {\bibfnamefont {J.~J.}\ \bibnamefont
  {Ramasco}}, \bibinfo {author} {\bibfnamefont {S.~N.}\ \bibnamefont
  {Dorogovtsev}}, \ and\ \bibinfo {author} {\bibfnamefont {R.}~\bibnamefont
  {Pastor-Satorras}},\ }\href@noop {} {\bibfield  {journal} {\bibinfo
  {journal} {Phys. Rev. E}\ }\textbf {\bibinfo {volume} {70}},\ \bibinfo
  {pages} {036106} (\bibinfo {year} {2004})}\BibitemShut {NoStop}%
\bibitem [{\citenamefont {Newman}\ \emph {et~al.}(2001)\citenamefont {Newman},
  \citenamefont {Strogatz},\ and\ \citenamefont {Watts}}]{newman2001random}%
  \BibitemOpen
  \bibfield  {author} {\bibinfo {author} {\bibfnamefont {M.}~\bibnamefont
  {Newman}}, \bibinfo {author} {\bibfnamefont {S.~H.}\ \bibnamefont
  {Strogatz}}, \ and\ \bibinfo {author} {\bibfnamefont {D.~J.}\ \bibnamefont
  {Watts}},\ }\href@noop {} {\bibfield  {journal} {\bibinfo  {journal} {Phys.
  Rev. E}\ }\textbf {\bibinfo {volume} {64}},\ \bibinfo {pages} {026118}
  (\bibinfo {year} {2001})}\BibitemShut {NoStop}%
\bibitem [{\citenamefont {Borgatti}\ and\ \citenamefont
  {Everett}(1997)}]{borgatti1997network}%
  \BibitemOpen
  \bibfield  {author} {\bibinfo {author} {\bibfnamefont {S.~P.}\ \bibnamefont
  {Borgatti}}\ and\ \bibinfo {author} {\bibfnamefont {M.~G.}\ \bibnamefont
  {Everett}},\ }\href@noop {} {\bibfield  {journal} {\bibinfo  {journal} {Soc.
  Networks}\ }\textbf {\bibinfo {volume} {19}},\ \bibinfo {pages} {243}
  (\bibinfo {year} {1997})}\BibitemShut {NoStop}%
\bibitem [{\citenamefont {Zhou}\ \emph {et~al.}(2007)\citenamefont {Zhou},
  \citenamefont {Ren}, \citenamefont {Medo},\ and\ \citenamefont
  {Zhang}}]{zhou2007bipartite}%
  \BibitemOpen
  \bibfield  {author} {\bibinfo {author} {\bibfnamefont {T.}~\bibnamefont
  {Zhou}}, \bibinfo {author} {\bibfnamefont {J.}~\bibnamefont {Ren}}, \bibinfo
  {author} {\bibfnamefont {M.}~\bibnamefont {Medo}}, \ and\ \bibinfo {author}
  {\bibfnamefont {Y.}~\bibnamefont {Zhang}},\ }\href@noop {} {\bibfield
  {journal} {\bibinfo  {journal} {Phys. Rev. E}\ }\textbf {\bibinfo {volume}
  {76}},\ \bibinfo {pages} {046115} (\bibinfo {year} {2007})}\BibitemShut
  {NoStop}%
\bibitem [{\citenamefont {Mukherjee}\ \emph {et~al.}(2011)\citenamefont
  {Mukherjee}, \citenamefont {Choudhury},\ and\ \citenamefont
  {Ganguly}}]{mukherjee2011understanding}%
  \BibitemOpen
  \bibfield  {author} {\bibinfo {author} {\bibfnamefont {A.}~\bibnamefont
  {Mukherjee}}, \bibinfo {author} {\bibfnamefont {M.}~\bibnamefont
  {Choudhury}}, \ and\ \bibinfo {author} {\bibfnamefont {N.}~\bibnamefont
  {Ganguly}},\ }\href@noop {} {\bibfield  {journal} {\bibinfo  {journal}
  {Physica A}\ }\textbf {\bibinfo {volume} {390}},\ \bibinfo {pages} {3602}
  (\bibinfo {year} {2011})}\BibitemShut {NoStop}%
\bibitem [{\citenamefont {Nacher}\ and\ \citenamefont
  {Akutsu}(2011)}]{nacher2011degree}%
  \BibitemOpen
  \bibfield  {author} {\bibinfo {author} {\bibfnamefont {J.~C.}\ \bibnamefont
  {Nacher}}\ and\ \bibinfo {author} {\bibfnamefont {T.}~\bibnamefont
  {Akutsu}},\ }\href@noop {} {\bibfield  {journal} {\bibinfo  {journal}
  {Physica A}\ }\textbf {\bibinfo {volume} {390}},\ \bibinfo {pages} {4636}
  (\bibinfo {year} {2011})}\BibitemShut {NoStop}%
\bibitem [{\citenamefont {Guillaume}\ and\ \citenamefont
  {Latapy}(2006)}]{guillaume2006bipartite}%
  \BibitemOpen
  \bibfield  {author} {\bibinfo {author} {\bibfnamefont {J.}~\bibnamefont
  {Guillaume}}\ and\ \bibinfo {author} {\bibfnamefont {M.}~\bibnamefont
  {Latapy}},\ }\href@noop {} {\bibfield  {journal} {\bibinfo  {journal}
  {Physica A}\ }\textbf {\bibinfo {volume} {371}},\ \bibinfo {pages} {795}
  (\bibinfo {year} {2006})}\BibitemShut {NoStop}%
\bibitem [{\citenamefont {Birmel{\'e}}(2009)}]{birmele2009scale}%
  \BibitemOpen
  \bibfield  {author} {\bibinfo {author} {\bibfnamefont {E.}~\bibnamefont
  {Birmel{\'e}}},\ }\href@noop {} {\bibfield  {journal} {\bibinfo  {journal}
  {Discrete Appl. Math.}\ }\textbf {\bibinfo {volume} {157}},\ \bibinfo {pages}
  {2267} (\bibinfo {year} {2009})}\BibitemShut {NoStop}%
\bibitem [{\citenamefont {Peruani}\ \emph {et~al.}(2007)\citenamefont
  {Peruani}, \citenamefont {Choudhury}, \citenamefont {Mukherjee},\ and\
  \citenamefont {Ganguly}}]{peruani2007emergence}%
  \BibitemOpen
  \bibfield  {author} {\bibinfo {author} {\bibfnamefont {F.}~\bibnamefont
  {Peruani}}, \bibinfo {author} {\bibfnamefont {M.}~\bibnamefont {Choudhury}},
  \bibinfo {author} {\bibfnamefont {A.}~\bibnamefont {Mukherjee}}, \ and\
  \bibinfo {author} {\bibfnamefont {N.}~\bibnamefont {Ganguly}},\ }\href@noop
  {} {\bibfield  {journal} {\bibinfo  {journal} {Europhys. Lett.}\ }\textbf
  {\bibinfo {volume} {79}},\ \bibinfo {pages} {28001} (\bibinfo {year}
  {2007})}\BibitemShut {NoStop}%
\bibitem [{\citenamefont {Choudhury}\ \emph {et~al.}(2010)\citenamefont
  {Choudhury}, \citenamefont {Ganguly}, \citenamefont {Maiti}, \citenamefont
  {Mukherjee}, \citenamefont {Brusch}, \citenamefont {Deutsch},\ and\
  \citenamefont {Peruani}}]{choudhury2010modeling}%
  \BibitemOpen
  \bibfield  {author} {\bibinfo {author} {\bibfnamefont {M.}~\bibnamefont
  {Choudhury}}, \bibinfo {author} {\bibfnamefont {N.}~\bibnamefont {Ganguly}},
  \bibinfo {author} {\bibfnamefont {A.}~\bibnamefont {Maiti}}, \bibinfo
  {author} {\bibfnamefont {A.}~\bibnamefont {Mukherjee}}, \bibinfo {author}
  {\bibfnamefont {L.}~\bibnamefont {Brusch}}, \bibinfo {author} {\bibfnamefont
  {A.}~\bibnamefont {Deutsch}}, \ and\ \bibinfo {author} {\bibfnamefont
  {F.}~\bibnamefont {Peruani}},\ }\href@noop {} {\bibfield  {journal} {\bibinfo
   {journal} {Phys. Rev. E}\ }\textbf {\bibinfo {volume} {81}},\ \bibinfo
  {pages} {036103} (\bibinfo {year} {2010})}\BibitemShut {NoStop}%
\bibitem [{\citenamefont {Newman}(2001{\natexlab{a}})}]{newman2001Structure}%
  \BibitemOpen
  \bibfield  {author} {\bibinfo {author} {\bibfnamefont {M.~E.}\ \bibnamefont
  {Newman}},\ }\href@noop {} {\bibfield  {journal} {\bibinfo  {journal} {Proc.
  Natl. Acad. Sci. USA}\ }\textbf {\bibinfo {volume} {98}},\ \bibinfo {pages}
  {404} (\bibinfo {year} {2001}{\natexlab{a}})}\BibitemShut {NoStop}%
\bibitem [{\citenamefont {Grossman}\ and\ \citenamefont
  {Ion}(1995)}]{grossman1995}%
  \BibitemOpen
  \bibfield  {author} {\bibinfo {author} {\bibfnamefont {J.~W.}\ \bibnamefont
  {Grossman}}\ and\ \bibinfo {author} {\bibfnamefont {P.~D.~F.}\ \bibnamefont
  {Ion}},\ }\href@noop {} {\bibfield  {journal} {\bibinfo  {journal}
  {Congressus Numerantium}\ } (\bibinfo {year} {1995})}\BibitemShut {NoStop}%
\bibitem [{\citenamefont {Watts}\ and\ \citenamefont
  {Strogatz}(1998)}]{watts1998collective}%
  \BibitemOpen
  \bibfield  {author} {\bibinfo {author} {\bibfnamefont {D.~J.}\ \bibnamefont
  {Watts}}\ and\ \bibinfo {author} {\bibfnamefont {S.~H.}\ \bibnamefont
  {Strogatz}},\ }\href@noop {} {\bibfield  {journal} {\bibinfo  {journal}
  {Nature}\ }\textbf {\bibinfo {volume} {393}},\ \bibinfo {pages} {440}
  (\bibinfo {year} {1998})}\BibitemShut {NoStop}%
\bibitem [{\citenamefont {Amaral}\ \emph {et~al.}(2000)\citenamefont {Amaral},
  \citenamefont {Scala}, \citenamefont {Barthelemy},\ and\ \citenamefont
  {Stanley}}]{Amaral2000Smallworld}%
  \BibitemOpen
  \bibfield  {author} {\bibinfo {author} {\bibfnamefont {L.~A.~N.}\
  \bibnamefont {Amaral}}, \bibinfo {author} {\bibfnamefont {A.}~\bibnamefont
  {Scala}}, \bibinfo {author} {\bibfnamefont {M.}~\bibnamefont {Barthelemy}}, \
  and\ \bibinfo {author} {\bibfnamefont {H.~E.}\ \bibnamefont {Stanley}},\
  }\href@noop {} {\bibfield  {journal} {\bibinfo  {journal} {Proc. Natl. Acad.
  Sci. USA}\ }\textbf {\bibinfo {volume} {97}},\ \bibinfo {pages} {11149}
  (\bibinfo {year} {2000})}\BibitemShut {NoStop}%
\bibitem [{\citenamefont {Newman}(2003{\natexlab{a}})}]{newman2003Function}%
  \BibitemOpen
  \bibfield  {author} {\bibinfo {author} {\bibfnamefont {M.~E.~J.}\
  \bibnamefont {Newman}},\ }\href@noop {} {\bibfield  {journal} {\bibinfo
  {journal} {SIAM Review}\ }\textbf {\bibinfo {volume} {45}},\ \bibinfo {pages}
  {167} (\bibinfo {year} {2003}{\natexlab{a}})}\BibitemShut {NoStop}%
\bibitem [{\citenamefont {Newman}\ \emph {et~al.}(2002)\citenamefont {Newman},
  \citenamefont {Forrest},\ and\ \citenamefont {Balthrop}}]{newman2002Email}%
  \BibitemOpen
  \bibfield  {author} {\bibinfo {author} {\bibfnamefont {M.~E.~J.}\
  \bibnamefont {Newman}}, \bibinfo {author} {\bibfnamefont {S.}~\bibnamefont
  {Forrest}}, \ and\ \bibinfo {author} {\bibfnamefont {J.}~\bibnamefont
  {Balthrop}},\ }\href@noop {} {\bibfield  {journal} {\bibinfo  {journal}
  {Phys. Rev. E}\ }\textbf {\bibinfo {volume} {66}},\ \bibinfo {pages} {47}
  (\bibinfo {year} {2002})}\BibitemShut {NoStop}%
\bibitem [{\citenamefont {Albert}\ \emph {et~al.}(1999)\citenamefont {Albert},
  \citenamefont {Jeong},\ and\ \citenamefont {Barabasi}}]{Albert1999www}%
  \BibitemOpen
  \bibfield  {author} {\bibinfo {author} {\bibfnamefont {R.}~\bibnamefont
  {Albert}}, \bibinfo {author} {\bibfnamefont {H.}~\bibnamefont {Jeong}}, \
  and\ \bibinfo {author} {\bibfnamefont {A.-L.}\ \bibnamefont {Barabasi}},\
  }\href@noop {} {\bibfield  {journal} {\bibinfo  {journal} {Nature}\ }\textbf
  {\bibinfo {volume} {401}},\ \bibinfo {pages} {130} (\bibinfo {year}
  {1999})}\BibitemShut {NoStop}%
\bibitem [{\citenamefont {Jeong}\ \emph {et~al.}(2001)\citenamefont {Jeong},
  \citenamefont {Mason}, \citenamefont {Barabasi},\ and\ \citenamefont
  {Oltvai}}]{jeong2001lethality}%
  \BibitemOpen
  \bibfield  {author} {\bibinfo {author} {\bibfnamefont {H.}~\bibnamefont
  {Jeong}}, \bibinfo {author} {\bibfnamefont {S.~P.}\ \bibnamefont {Mason}},
  \bibinfo {author} {\bibfnamefont {A.-L.}\ \bibnamefont {Barabasi}}, \ and\
  \bibinfo {author} {\bibfnamefont {Z.~N.}\ \bibnamefont {Oltvai}},\
  }\href@noop {} {\bibfield  {journal} {\bibinfo  {journal} {Nature}\ }\textbf
  {\bibinfo {volume} {411}},\ \bibinfo {pages} {41} (\bibinfo {year}
  {2001})}\BibitemShut {NoStop}%
\bibitem [{\citenamefont {Martinez}(2018)}]{Martinez1991foodweb}%
  \BibitemOpen
  \bibfield  {author} {\bibinfo {author} {\bibfnamefont {N.~D.}\ \bibnamefont
  {Martinez}},\ }\href@noop {} {\bibfield  {journal} {\bibinfo  {journal}
  {Ecol. Monogr.}\ }\textbf {\bibinfo {volume} {61}},\ \bibinfo {pages} {367}
  (\bibinfo {year} {2018})}\BibitemShut {NoStop}%
\bibitem [{\citenamefont {Newman}(2002)}]{newman2002Assortative}%
  \BibitemOpen
  \bibfield  {author} {\bibinfo {author} {\bibfnamefont {M.~E.~J.}\
  \bibnamefont {Newman}},\ }\href@noop {} {\bibfield  {journal} {\bibinfo
  {journal} {Phys. Rev. Lett.}\ }\textbf {\bibinfo {volume} {89}},\ \bibinfo
  {pages} {129} (\bibinfo {year} {2002})}\BibitemShut {NoStop}%
\bibitem [{\citenamefont {Newman}(2003{\natexlab{b}})}]{newman2003Mixing}%
  \BibitemOpen
  \bibfield  {author} {\bibinfo {author} {\bibfnamefont {M.~E.~J.}\
  \bibnamefont {Newman}},\ }\href@noop {} {\bibfield  {journal} {\bibinfo
  {journal} {Phys. Rev. E}\ }\textbf {\bibinfo {volume} {67}},\ \bibinfo
  {pages} {026126} (\bibinfo {year} {2003}{\natexlab{b}})}\BibitemShut
  {NoStop}%
\bibitem [{\citenamefont {Newman}\ and\ \citenamefont
  {Park}(2003)}]{newman2003Social}%
  \BibitemOpen
  \bibfield  {author} {\bibinfo {author} {\bibfnamefont {M.~E.~J.}\
  \bibnamefont {Newman}}\ and\ \bibinfo {author} {\bibfnamefont
  {J.}~\bibnamefont {Park}},\ }\href@noop {} {\bibfield  {journal} {\bibinfo
  {journal} {Phys. Rev. E}\ }\textbf {\bibinfo {volume} {68}},\ \bibinfo
  {pages} {036122} (\bibinfo {year} {2003})}\BibitemShut {NoStop}%
\bibitem [{\citenamefont {Wilf}(2013)}]{wilf2013generatingfunctionology}%
  \BibitemOpen
  \bibfield  {author} {\bibinfo {author} {\bibfnamefont {H.~S.}\ \bibnamefont
  {Wilf}},\ }\href@noop {} {\emph {\bibinfo {title}
  {Generatingfunctionology}}}\ (\bibinfo  {publisher} {Elsevier},\ \bibinfo
  {year} {2013})\BibitemShut {NoStop}%
\bibitem [{\citenamefont {Pemmaraju}\ and\ \citenamefont
  {Skiena}(2003)}]{pemmaraju2003computational}%
  \BibitemOpen
  \bibfield  {author} {\bibinfo {author} {\bibfnamefont {S.}~\bibnamefont
  {Pemmaraju}}\ and\ \bibinfo {author} {\bibfnamefont {S.}~\bibnamefont
  {Skiena}},\ }\href@noop {} {\emph {\bibinfo {title} {Computational Discrete
  Mathematics: Combinatorics and Graph Theory with
  Mathematica\textsuperscript{\textregistered}}}}\ (\bibinfo  {publisher}
  {Cambridge University Press},\ \bibinfo {address} {Cambridge},\ \bibinfo
  {year} {2003})\BibitemShut {NoStop}%
\bibitem [{\citenamefont {Souma}\ \emph {et~al.}(2003)\citenamefont {Souma},
  \citenamefont {Fujiwara},\ and\ \citenamefont {Aoyama}}]{souma2003complex}%
  \BibitemOpen
  \bibfield  {author} {\bibinfo {author} {\bibfnamefont {W.}~\bibnamefont
  {Souma}}, \bibinfo {author} {\bibfnamefont {Y.}~\bibnamefont {Fujiwara}}, \
  and\ \bibinfo {author} {\bibfnamefont {H.}~\bibnamefont {Aoyama}},\
  }\href@noop {} {\bibfield  {journal} {\bibinfo  {journal} {Physica A}\
  }\textbf {\bibinfo {volume} {324}},\ \bibinfo {pages} {396} (\bibinfo {year}
  {2003})}\BibitemShut {NoStop}%
\bibitem [{\citenamefont {Newman}(2001{\natexlab{b}})}]{newman2001scientific}%
  \BibitemOpen
  \bibfield  {author} {\bibinfo {author} {\bibfnamefont {M.~E.~J.}\
  \bibnamefont {Newman}},\ }\href@noop {} {\bibfield  {journal} {\bibinfo
  {journal} {Phys. Rev. E}\ }\textbf {\bibinfo {volume} {64}},\ \bibinfo
  {pages} {016131} (\bibinfo {year} {2001}{\natexlab{b}})}\BibitemShut
  {NoStop}%
\bibitem [{\citenamefont {Onody}\ and\ \citenamefont
  {de~Castro}(2004)}]{onody2004complex}%
  \BibitemOpen
  \bibfield  {author} {\bibinfo {author} {\bibfnamefont {R.~N.}\ \bibnamefont
  {Onody}}\ and\ \bibinfo {author} {\bibfnamefont {P.}~\bibnamefont
  {de~Castro}},\ }\href@noop {} {\bibfield  {journal} {\bibinfo  {journal}
  {Phys. Rev. E}\ }\textbf {\bibinfo {volume} {70}},\ \bibinfo {pages} {037103}
  (\bibinfo {year} {2004})}\BibitemShut {NoStop}%
\bibitem [{\citenamefont {Gibbons}(1985)}]{gibbons1985algorithmic}%
  \BibitemOpen
  \bibfield  {author} {\bibinfo {author} {\bibfnamefont {A.}~\bibnamefont
  {Gibbons}},\ }\href@noop {} {\emph {\bibinfo {title} {Algorithmic Graph
  Theory}}}\ (\bibinfo  {publisher} {Cambridge University Press},\ \bibinfo
  {address} {Cambridge},\ \bibinfo {year} {1985})\BibitemShut {NoStop}%
\bibitem [{\citenamefont {West}\ \emph {et~al.}(2001)\citenamefont {West} \emph
  {et~al.}}]{west2001introduction}%
  \BibitemOpen
  \bibfield  {author} {\bibinfo {author} {\bibfnamefont {D.~B.}\ \bibnamefont
  {West}} \emph {et~al.},\ }\href@noop {} {\emph {\bibinfo {title}
  {Introduction to Graph Theory}}},\ Vol.~\bibinfo {volume} {2}\ (\bibinfo
  {publisher} {Prentice Hall},\ \bibinfo {address} {Upper Saddle River},\
  \bibinfo {year} {2001})\BibitemShut {NoStop}%
\bibitem [{\citenamefont {Gross}\ and\ \citenamefont
  {Yellen}(2005)}]{gross2005graph}%
  \BibitemOpen
  \bibfield  {author} {\bibinfo {author} {\bibfnamefont {J.~L.}\ \bibnamefont
  {Gross}}\ and\ \bibinfo {author} {\bibfnamefont {J.}~\bibnamefont {Yellen}},\
  }\href@noop {} {\emph {\bibinfo {title} {Graph Theory and its
  Applications}}}\ (\bibinfo  {publisher} {CRC press},\ \bibinfo {year}
  {2005})\BibitemShut {NoStop}%
\bibitem [{\citenamefont {Bollob{\'a}s}(2013)}]{bollobas2013modern}%
  \BibitemOpen
  \bibfield  {author} {\bibinfo {author} {\bibfnamefont {B.}~\bibnamefont
  {Bollob{\'a}s}},\ }\href@noop {} {\emph {\bibinfo {title} {Modern Graph
  Theory}}},\ Vol.\ \bibinfo {volume} {184}\ (\bibinfo  {publisher} {Springer
  Science \& Business Media},\ \bibinfo {year} {2013})\BibitemShut {NoStop}%
\bibitem [{\citenamefont {Granovetter}(1973)}]{granovetter1973strength}%
  \BibitemOpen
  \bibfield  {author} {\bibinfo {author} {\bibfnamefont {M.~S.}\ \bibnamefont
  {Granovetter}},\ }\href@noop {} {\bibfield  {journal} {\bibinfo  {journal}
  {Am. J. Sociol.}\ }\textbf {\bibinfo {volume} {78}},\ \bibinfo {pages} {1360}
  (\bibinfo {year} {1973})}\BibitemShut {NoStop}%
\bibitem [{\citenamefont {Clark}\ and\ \citenamefont
  {Holton}(1991)}]{clark1991first}%
  \BibitemOpen
  \bibfield  {author} {\bibinfo {author} {\bibfnamefont {J.}~\bibnamefont
  {Clark}}\ and\ \bibinfo {author} {\bibfnamefont {D.~A.}\ \bibnamefont
  {Holton}},\ }\href@noop {} {\emph {\bibinfo {title} {A First Look at Graph
  Theory}}}\ (\bibinfo  {publisher} {World Scientific Publishing Co Inc},\
  \bibinfo {year} {1991})\BibitemShut {NoStop}%
\bibitem [{\citenamefont {Newman}(2001{\natexlab{c}})}]{newman2001scientific2}%
  \BibitemOpen
  \bibfield  {author} {\bibinfo {author} {\bibfnamefont {M.~E.~J.}\
  \bibnamefont {Newman}},\ }\href@noop {} {\bibfield  {journal} {\bibinfo
  {journal} {Phys. Rev. E}\ }\textbf {\bibinfo {volume} {64}},\ \bibinfo
  {pages} {016132} (\bibinfo {year} {2001}{\natexlab{c}})}\BibitemShut
  {NoStop}%
\bibitem [{\citenamefont {Luczkovich}\ \emph {et~al.}(2003)\citenamefont
  {Luczkovich}, \citenamefont {Borgatti}, \citenamefont {Johnson},\ and\
  \citenamefont {Everett}}]{luczkovich2003defining}%
  \BibitemOpen
  \bibfield  {author} {\bibinfo {author} {\bibfnamefont {J.~J.}\ \bibnamefont
  {Luczkovich}}, \bibinfo {author} {\bibfnamefont {S.~P.}\ \bibnamefont
  {Borgatti}}, \bibinfo {author} {\bibfnamefont {J.~C.}\ \bibnamefont
  {Johnson}}, \ and\ \bibinfo {author} {\bibfnamefont {M.~G.}\ \bibnamefont
  {Everett}},\ }\href@noop {} {\bibfield  {journal} {\bibinfo  {journal} {J.
  Theor. Biol.}\ }\textbf {\bibinfo {volume} {220}},\ \bibinfo {pages} {303}
  (\bibinfo {year} {2003})}\BibitemShut {NoStop}%
\bibitem [{\citenamefont {Pastor-Satorras}\ and\ \citenamefont
  {Vespignani}(2007)}]{pastor2007evolution}%
  \BibitemOpen
  \bibfield  {author} {\bibinfo {author} {\bibfnamefont {R.}~\bibnamefont
  {Pastor-Satorras}}\ and\ \bibinfo {author} {\bibfnamefont {A.}~\bibnamefont
  {Vespignani}},\ }\href@noop {} {\emph {\bibinfo {title} {Evolution and
  Structure of the Internet: A Statistical Physics Approach}}}\ (\bibinfo
  {publisher} {Cambridge University Press},\ \bibinfo {address} {Cambridge},\
  \bibinfo {year} {2007})\BibitemShut {NoStop}%
\bibitem [{\citenamefont {Barrat}\ \emph {et~al.}(2004)\citenamefont {Barrat},
  \citenamefont {Barthelemy}, \citenamefont {Pastor-Satorras},\ and\
  \citenamefont {Vespignani}}]{barrat2004architecture}%
  \BibitemOpen
  \bibfield  {author} {\bibinfo {author} {\bibfnamefont {A.}~\bibnamefont
  {Barrat}}, \bibinfo {author} {\bibfnamefont {M.}~\bibnamefont {Barthelemy}},
  \bibinfo {author} {\bibfnamefont {R.}~\bibnamefont {Pastor-Satorras}}, \ and\
  \bibinfo {author} {\bibfnamefont {A.}~\bibnamefont {Vespignani}},\
  }\href@noop {} {\bibfield  {journal} {\bibinfo  {journal} {Proc. Natl. Acad.
  Sci. USA}\ }\textbf {\bibinfo {volume} {101}},\ \bibinfo {pages} {3747}
  (\bibinfo {year} {2004})}\BibitemShut {NoStop}%
\bibitem [{\citenamefont {Li}\ and\ \citenamefont
  {Wang}(2015)}]{li2015generating}%
  \BibitemOpen
  \bibfield  {author} {\bibinfo {author} {\bibfnamefont {M.}~\bibnamefont
  {Li}}\ and\ \bibinfo {author} {\bibfnamefont {B.}~\bibnamefont {Wang}},\ }in\
  \href@noop {} {\emph {\bibinfo {booktitle} {J. Phys. Conf. Ser.}}},\ Vol.\
  \bibinfo {volume} {604}\ (\bibinfo {organization} {IOP Publishing},\ \bibinfo
  {year} {2015})\ p.\ \bibinfo {pages} {012013}\BibitemShut {NoStop}%
\bibitem [{\citenamefont {Miller}(2015)}]{miller2015probability}%
  \BibitemOpen
  \bibfield  {author} {\bibinfo {author} {\bibfnamefont {S.~J.}\ \bibnamefont
  {Miller}},\ }\href@noop {} {\emph {\bibinfo {title} {The Probability
  Lifesaver}}}\ (\bibinfo  {publisher} {Princeton University Press},\ \bibinfo
  {address} {Princeton, NJ},\ \bibinfo {year} {2015})\BibitemShut {NoStop}%
\bibitem [{\citenamefont {Bender}\ and\ \citenamefont
  {Canfield}(1978)}]{bender1978asymptotic}%
  \BibitemOpen
  \bibfield  {author} {\bibinfo {author} {\bibfnamefont {E.~A.}\ \bibnamefont
  {Bender}}\ and\ \bibinfo {author} {\bibfnamefont {E.~R.}\ \bibnamefont
  {Canfield}},\ }\href@noop {} {\bibfield  {journal} {\bibinfo  {journal} {J.
  Comb. Theory A}\ }\textbf {\bibinfo {volume} {24}},\ \bibinfo {pages} {296}
  (\bibinfo {year} {1978})}\BibitemShut {NoStop}%
\bibitem [{\citenamefont {Bollob{\'a}s}(1980)}]{bollobas1980probabilistic}%
  \BibitemOpen
  \bibfield  {author} {\bibinfo {author} {\bibfnamefont {B.}~\bibnamefont
  {Bollob{\'a}s}},\ }\href@noop {} {\bibfield  {journal} {\bibinfo  {journal}
  {Eur. J. Combin.}\ }\textbf {\bibinfo {volume} {1}},\ \bibinfo {pages} {311}
  (\bibinfo {year} {1980})}\BibitemShut {NoStop}%
\bibitem [{\citenamefont {Hidalgo}\ and\ \citenamefont
  {Hausmann}(2009)}]{hidalgo2009building}%
  \BibitemOpen
  \bibfield  {author} {\bibinfo {author} {\bibfnamefont {C.~A.}\ \bibnamefont
  {Hidalgo}}\ and\ \bibinfo {author} {\bibfnamefont {R.}~\bibnamefont
  {Hausmann}},\ }\href@noop {} {\bibfield  {journal} {\bibinfo  {journal}
  {Proc. Natl. Acad. Sci. USA}\ }\textbf {\bibinfo {volume} {106}},\ \bibinfo
  {pages} {10570} (\bibinfo {year} {2009})}\BibitemShut {NoStop}%
\bibitem [{\citenamefont {Battiston}\ and\ \citenamefont
  {Catanzaro}(2004)}]{battiston2004statistical}%
  \BibitemOpen
  \bibfield  {author} {\bibinfo {author} {\bibfnamefont {S.}~\bibnamefont
  {Battiston}}\ and\ \bibinfo {author} {\bibfnamefont {M.}~\bibnamefont
  {Catanzaro}},\ }\href@noop {} {\bibfield  {journal} {\bibinfo  {journal}
  {Eur. Phys. J. B}\ }\textbf {\bibinfo {volume} {38}},\ \bibinfo {pages} {345}
  (\bibinfo {year} {2004})}\BibitemShut {NoStop}%
\bibitem [{\citenamefont {Abbasi}\ \emph {et~al.}(2011)\citenamefont {Abbasi},
  \citenamefont {Hossain}, \citenamefont {Uddin},\ and\ \citenamefont
  {Rasmussen}}]{abbasi2011evolutionary}%
  \BibitemOpen
  \bibfield  {author} {\bibinfo {author} {\bibfnamefont {A.}~\bibnamefont
  {Abbasi}}, \bibinfo {author} {\bibfnamefont {L.}~\bibnamefont {Hossain}},
  \bibinfo {author} {\bibfnamefont {S.}~\bibnamefont {Uddin}}, \ and\ \bibinfo
  {author} {\bibfnamefont {K.~J.}\ \bibnamefont {Rasmussen}},\ }\href@noop {}
  {\bibfield  {journal} {\bibinfo  {journal} {Scientometrics}\ }\textbf
  {\bibinfo {volume} {89}},\ \bibinfo {pages} {687} (\bibinfo {year}
  {2011})}\BibitemShut {NoStop}%
\bibitem [{\citenamefont {Chakraborty}\ and\ \citenamefont
  {Manna}(2010)}]{chakraborty2010weighted}%
  \BibitemOpen
  \bibfield  {author} {\bibinfo {author} {\bibfnamefont {A.}~\bibnamefont
  {Chakraborty}}\ and\ \bibinfo {author} {\bibfnamefont {S.}~\bibnamefont
  {Manna}},\ }\href@noop {} {\bibfield  {journal} {\bibinfo  {journal} {Phys.
  Rev. E}\ }\textbf {\bibinfo {volume} {81}},\ \bibinfo {pages} {016111}
  (\bibinfo {year} {2010})}\BibitemShut {NoStop}%
\bibitem [{\citenamefont {Newman}(2006)}]{newman2006finding}%
  \BibitemOpen
  \bibfield  {author} {\bibinfo {author} {\bibfnamefont {M.~E.}\ \bibnamefont
  {Newman}},\ }\href@noop {} {\bibfield  {journal} {\bibinfo  {journal} {Phys.
  Rev. E}\ }\textbf {\bibinfo {volume} {74}},\ \bibinfo {pages} {036104}
  (\bibinfo {year} {2006})}\BibitemShut {NoStop}%
\bibitem [{\citenamefont {Barber}(2007)}]{barber2007modularity}%
  \BibitemOpen
  \bibfield  {author} {\bibinfo {author} {\bibfnamefont {M.~J.}\ \bibnamefont
  {Barber}},\ }\href@noop {} {\bibfield  {journal} {\bibinfo  {journal} {Phys.
  Rev. E}\ }\textbf {\bibinfo {volume} {76}},\ \bibinfo {pages} {066102}
  (\bibinfo {year} {2007})}\BibitemShut {NoStop}%
\bibitem [{\citenamefont {Zhang}\ \emph {et~al.}(2008)\citenamefont {Zhang},
  \citenamefont {Wang}, \citenamefont {Li}, \citenamefont {Li}, \citenamefont
  {Di},\ and\ \citenamefont {Fan}}]{zhang2008clustering}%
  \BibitemOpen
  \bibfield  {author} {\bibinfo {author} {\bibfnamefont {P.}~\bibnamefont
  {Zhang}}, \bibinfo {author} {\bibfnamefont {J.}~\bibnamefont {Wang}},
  \bibinfo {author} {\bibfnamefont {X.}~\bibnamefont {Li}}, \bibinfo {author}
  {\bibfnamefont {M.}~\bibnamefont {Li}}, \bibinfo {author} {\bibfnamefont
  {Z.}~\bibnamefont {Di}}, \ and\ \bibinfo {author} {\bibfnamefont
  {Y.}~\bibnamefont {Fan}},\ }\href@noop {} {\bibfield  {journal} {\bibinfo
  {journal} {Physica A}\ }\textbf {\bibinfo {volume} {387}},\ \bibinfo {pages}
  {6869} (\bibinfo {year} {2008})}\BibitemShut {NoStop}%
\bibitem [{\citenamefont {Larremore}\ \emph {et~al.}(2014)\citenamefont
  {Larremore}, \citenamefont {Clauset},\ and\ \citenamefont
  {Jacobs}}]{larremore2014efficiently}%
  \BibitemOpen
  \bibfield  {author} {\bibinfo {author} {\bibfnamefont {D.~B.}\ \bibnamefont
  {Larremore}}, \bibinfo {author} {\bibfnamefont {A.}~\bibnamefont {Clauset}},
  \ and\ \bibinfo {author} {\bibfnamefont {A.~Z.}\ \bibnamefont {Jacobs}},\
  }\href@noop {} {\bibfield  {journal} {\bibinfo  {journal} {Phys. Rev. E}\
  }\textbf {\bibinfo {volume} {90}},\ \bibinfo {pages} {012805} (\bibinfo
  {year} {2014})}\BibitemShut {NoStop}%
\end{thebibliography}%

\end{document}